\documentclass[manuscript]{acmart}
\usepackage{amsfonts}
\usepackage{amsmath}
\usepackage{amsthm}
\usepackage{ulem}
\usepackage{cancel}
\usepackage{comment}
\usepackage{bm}
\usepackage{tikz}
\usepackage{algorithm}
\usepackage{algorithmicx}
\usepackage{algpseudocode}
\usepackage{graphicx}
\usetikzlibrary{calc}
\usetikzlibrary{intersections}
\usetikzlibrary{shapes}
\usetikzlibrary{shapes.multipart}
\usetikzlibrary{shapes.geometric}
\usetikzlibrary{arrows}
\usepackage{pgf}
\usepackage{pgfplots}
\pgfplotsset{compat=1.18}
\usepackage{pgfplotstable}
\usepgfplotslibrary{fillbetween}
\usepackage{tikz-3dplot}
\usetikzlibrary{3d}
\usepackage{wrapfig}
\usepackage{subcaption}
\usepackage{multirow}
\usepackage{subfiles}
\usepackage{lineno}
\usepackage{hyperref}
\usepackage{enumitem}
\usepackage{makecell}
\usepackage[vskip=1pt,leftmargin=2em]{quoting}
\usepackage{adjustbox}

\usepackage{xspace}
\usepackage{booktabs}
\usepackage{subcaption}
\usepackage[mode=buildnew]{standalone}

\algrenewcommand\algorithmicindent{0.5em}

\newcommand{\matt}[1]{({\color{green} \bf Matt: #1})}

\newcommand{\nusrat}[1]{({\color{blue} Nusrat: \bf #1})}
\newcommand{\ft}[1]{{\color{red}\{FT: #1\}}}
\newcommand{\ignore}[1]{}

\newcommand{\torepo}[1]{}

\tikzset{
   operator/.style = {rectangle, thick, draw, rounded corners, minimum width=0.6cm, minimum height = 1cm},
   choice/.style = {diamond, thick, draw, inner sep=0},
   sgvertex/.style = {draw, circle, minimum width=10mm, thick}
}

\newtheorem{definition}{Definition}

\renewcommand{\sectionautorefname}{\S\kern-2pt}

\newcommand{\X}{\mathcal{X}}  

\newcommand{\pre}[1]{{\phi^{#1}_{\mathcal{X}}}}
\newcommand{\post}[1]{{\phi^{#1}_{\mathcal{Y}}}}

\newcommand{\DNN}{N} 
\newcommand{\Encoder}{\mathcal{E}}
\newcommand{\Decoder}{\mathcal{D}}
\newcommand{\R}{\mathbb{R}}  
\newcommand{\Normal}{\mathcal{N}(0, 1)}
\newcommand{\technique}{\textsc{rbt4dnn}\xspace}

\renewcommand{\rq}[1]{\textbf{RQ#1}}

\title{\technique: Requirements-based Testing of Neural Networks}

\author{Nusrat Jahan Mozumder}
\email{nm8tm@virginia.edu}
\orcid{0009-0003-1802-5150}
\affiliation{%
  \institution{Department of Computer Science, University of Virginia}
  \city{Charlottesville}
  \state{Virginia}
  \country{USA}
}

\author{Felipe Toledo}
\email{ft8bn@virginia.edu}
\orcid{0000-0002-5632-7518}
\affiliation{%
  \institution{Department of Computer Science, University of Virginia}
  \city{Charlottesville}
  \state{Virginia}
  \country{USA}
}

\author{Swaroopa Dola}
\email{sd4tx@virginia.edu}
\orcid{0000-0002-9831-2744}
\affiliation{%
  \institution{Department of Computer Engineering, University of Virginia}
  \city{Charlottesville}
  \state{Virginia}
  \country{USA}
}

\author{Matthew B. Dwyer}
\email{matthewbdwyer@virginia.edu}
\orcid{0000-0002-1937-1544}
\affiliation{%
  \institution{Department of Computer Science, University of Virginia}
  \city{Charlottesville}
  \state{Virginia}
  \country{USA}
}

\begin{document}

\begin{abstract}
Testing allows developers to determine whether a system functions as expected. When such systems include deep neural networks (DNNs),
Testing becomes challenging, as DNNs approximate functions for which the formalization of functional requirements is intractable. This prevents the application of well-developed approaches to requirements-based testing to DNNs.

To address this, we propose a requirements-based testing method (\technique) that uses natural language requirements statements. 
These statements use a glossary of terms to define a semantic feature space that can be leveraged for test input generation. \technique formalizes preconditions of functional requirements as logical combinations of those semantic features. Training data matching these feature combinations can be used to fine-tune a generative model to reliably produce test inputs satisfying the precondition. Executing these tests on a trained DNN enables comparing its output to the expected requirement postcondition behavior. We propose two use cases for \technique: (1) given requirements defining DNN correctness properties, \technique comprises a novel approach for detecting faults, and (2) during development, requirements-guided exploration of model behavior can provide developers with feedback on model generalization. Our further evaluation shows that \technique-generated tests are realistic, diverse, and aligned with requirement preconditions, enabling targeted analysis of model behavior and effective fault detection.

\end{abstract}



\maketitle

\section{Introduction}
\label{sec:intro}

As deep neural networks (DNN) become increasingly capable the breadth of systems that will include them as components will grow.  
A key question for system developers is how to 
determine whether a trained DNN is \textit{fit} for 
inclusion in a system.   
The most common approach to answering this question
is to evaluate \textit{test accuracy}.
In this approach, a \textit{test set} is defined that consists of input samples
that are each paired with a specific expected
DNN output -- for categorical DNNs the expected output is called a \textit{test label}.
Test accuracy is the percentage of input samples in the test set
for which the DNN produces the expected output associated with the input~\cite{goodfellow2016deep}.

This approach resembles a common strategy used in traditional software testing where
the expected outputs define an input-specific \textit{test oracle}~\cite{barr2014oracle}.
While this type of DNN testing may be appropriate for
some DNNs, and systems that include them, it is limited in several ways.
First, such test oracles are specific to individual inputs, which means that
they cannot be applied to other inputs and this \textbf{limits the ability to test how well a DNN generalizes over its input domain}.
Second, such test oracles are very narrowly defined, e.g., a single label for
categorical networks, which \textbf{limits the ability to capture broader definitions of
acceptable model behavior}.
Third, the test set is typically a randomly chosen held out sample of labeled training data, which \textbf{limits
the ability of testing to focus on classes of inputs that might be of interest
to  developers}.

There is a rich literature on requirements-based testing for traditional software systems that would seem to address these limitations~\cite{unterkalmsteiner2014taxonomy}.
A common focus of these methods is on \textit{functional requirements}, e.g.,~\cite{uusitalo2011structured,veizaga2021systematically},
which define a \textit{precondition} -- defining 
a class of inputs -- and a \textit{postcondition} -- defining
a general oracle that is applicable across that class of inputs.
This type of testing is standard for many critical systems, such as
aircraft or medical devices, precisely because it explicitly relates tested behavior to stated functional requirements~\cite{cleland2014software}.
Not only should testing be related to requirements, but it should be thorough in exercising behavior related to those requirements.  
To address this need, researchers have developed frameworks 
that analyze formalizations of functional requirements to generate test inputs that thoroughly
cover the preconditions and then check the associated postconditions~\cite{rayadurgam2001test,nebut2006automatic,whalen2006coverage,pecheur2009formal}.

It is, however, challenging to directly apply existing research on requirements-based testing to systems that include machine \textit{learned components} (LC), because it can be challenging
to formalize LC requirements.
Consider the safe driving requirement expressed in
\S46.2-816 of the Virginia Driving Code -- which states that a vehicle should not follow another too closely~\cite{CodeOfVirginia} where
``closely'' is
defined using the 2, 3, or 4-second rule~\cite{VirginiaDriversManual}[Section 3].
Imagine an autonomous driving system that incorporates an LC
that accepts camera inputs and produces outputs that direct
the acceleration and steering of the vehicle.
To levy this requirement on such an autonomous driving system,
and the LC within it, 
one might begin by expressing a necessary condition for
the requirement in natural language as:
\begin{quoting}
\textit{If} \textbf{a vehicle is within 10 meters, in front, and in the same lane}\textit{, then the LC shall} \textbf{not accelerate.}             
\end{quoting}
The intent of this necessary condition is to prevent the autonomous
vehicle from becoming too close to another vehicle.
This informally stated functional requirement has a precondition, between the \textit{If} and \textit{then}, 
that defines features of the scene that must be present for the requirement to be active, and
a postcondition, after the \textit{shall}, that defines the vehicle control
actions that are permitted when the requirement is active, e.g., that acceleration is $\le 0$.

To apply existing requirements-based testing approaches one needs to formalize this requirement.
To do this, one would need represent the features mentioned in the precondition
within the space of 3-channel 900 by 256 pixel images flowing to the LC from a camera
mounted on the windshield of the car -- which we refer to as the \textit{ego} vehicle.
Precisely representing a \textit{semantic feature} like  \textbf{a vehicle is in the same lane} in the pixel space is, however,
extremely challenging given variability in
the type of vehicle, its color, its relative position to the ego vehicle, its relative position within the
lane, lane curvature, the natural variability in lane markings, and myriad other factors related to lighting and image quality.  
Unfortunately, despite longstanding recognition of the need to support such semantic features, e.g., ~\cite{seshia2018formal,rahimi2019toward,vogelsang2019requirements}, and despite efforts towards that end, ~\cite{hu2022if,hu2022check,toledo2023deeper}, there are no broadly applicable semantic feature-based formal specification approaches for LCs.

In this paper, we present a method for \textit{requirements-based testing of deep neural networks} (\technique) that leverages natural language statements of functional requirements expressed over semantic features.
Our key insight is that we can side-step the challenge of formalizing
such requirements by leveraging the fact that the latent-space of
modern generative models comprises a representation of semantic features~\cite{kwon2023diffusion}.
Unfortunately, such a latent-space is uninterpretable and searching for a specific semantic feature is intractable, so
precisely locating where the combination of semantic features
in a requirement precondition resides in the latent space is challenging.
\technique side-steps that challenge by 
fine-tuning the pre-trained generative model using a \text{low-rank adaptation} (LoRA)~\cite{hu2022lora} with a small number of input samples that share the semantic
features of the precondition.  The resulting LoRA learns 
 the combination of semantic features in the inputs upon which it was trained and can then
 be used to generate novel inputs that share that feature combination.
\technique can then run the LC under test on generated inputs and check the output 
on the postcondition, e.g., acceleration $\le 0$, associated with the logical combination 
of semantic
feature combinations present in the precondition, e.g., \textbf{within 10 meters}
and \textbf{in front} and \textbf{in the same lane}.

\paragraph{An Illustrative Example}
As in prior work on formal requirements testing, ~\cite{rayadurgam2001test,nebut2006automatic,whalen2006coverage,pecheur2009formal},
we focus on necessary conditions for correctness.
For example, rather than attempt to encode the complete details of \S46.2-816, the above requirement
uses a precondition that is more restrictive than the law -- 10 meters is less than
the closest distance defined by the 2, 3, or 4-second rule -- and a postcondition that
is more liberal than the law requires -- it precludes acceleration, but does not require braking.
This ensures that violations of the requirement
are violations of \S46.2-816, so that requirement violations provide feedback to 
LC developers about behavior that violates the legal requirement.

\begin{figure}[t]
    \centering
    \includegraphics[width=0.58\textwidth]{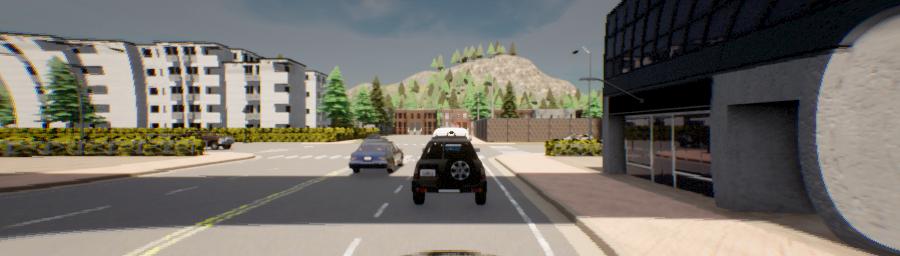}%
    \hfill
    \includegraphics[width=0.4\textwidth]{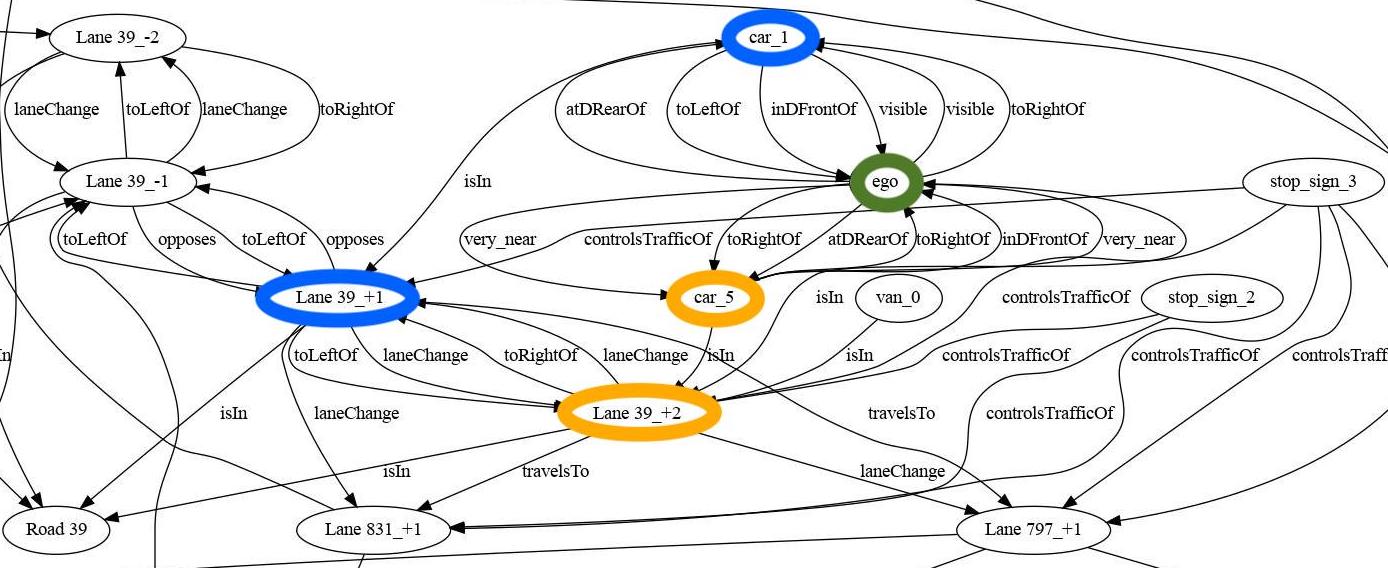} \\ 
    {\small
    Ego vehicle is \textcolor{green}{in the rightmost lane}. \\
    A car is \textcolor{red}{7 to 10 meters away}, \textcolor{orange}{in front}, and \textcolor{yellow}{in the same lane}. \\
    A car is \textcolor{blue}{16 to 25 meters away}, \textcolor{orange}{in front}, and \textcolor{violet}{in the lane to the left}.}
    \vspace{-2mm}
    \caption{Camera image input (top left) and fragment of the associated scene graph (top right) with 3 statements comprised of glossary terms describing relationships between ego and 
    elements of the scene (bottom).
}
    \label{fig:image-sg-terms}
\end{figure}

We focus on functional requirements that express necessary conditions on the input-output
relation of the LC~\cite{seshia2018formal} and support requirements that are expressed using \textit{structured natural language} (SNL)~\cite{mavin2009easy,uusitalo2011structured,veizaga2021systematically,grosser2023comparative}.To align with standards in requirements engineering, 
we assume that  preconditions are expressed using a
set of predefined \textit{glossary terms} where each term corresponds to 
a domain-specific semantic feature~\cite{lamsweerde2009requirements,pohl2010requirements,arora2016automated}.
To illustrate, consider the camera image on the top left of Figure~\ref{fig:image-sg-terms}
which satisfies the precondition of the driving requirement stated above -- \textbf{a vehicle is
within 10 meters, in front, and in the same lane}.
Here the glossary terms capture information about the presence of, distance to, direction to,
and lane occupancy of vehicles relative to the ego.
The bottom of Figure~\ref{fig:image-sg-terms} shows 
6 distinct glossary terms -- in different colors, where combinations of terms are used to define an entity in the scene.

Next, \technique maps the training data inputs to the set of glossary terms.
It is well-understood that this type of data labeling can be expensive so
\technique leverages various forms
of auto-labeling~\cite{das2020goggles,ratner2020snorkel} that use algorithmic
or machine learning techniques to convert image inputs to glossary terms.
For example, the upper right of Figure~\ref{fig:image-sg-terms} shows a fragment
of a scene graph (SG) that can be generated using computer-vision based ML 
models, e.g.,~\cite{malawade2022roadscene2vec,li2024pixels}.
SG vertices encode a road's structural elements,
the vehicles in the image, and other features like stop signs and
traffic lights.
In the graph, we highlight the vertices encoding the ego vehicle (green), two cars (blue, gold), and two lanes (blue gold).
The edges in the graph encode semantic relations in the scene, e.g., that ego \textbf{isIn} the
lane in gold (39\_+2).  
Glossary term labeling can be formulated as a search rooted at the ego vehicle, where paths
in the search are translated to semantic feature labels.
This labeling approach is appropriate for the autonomous driving dataset, but we 
present more general methods in Section~\ref{sec:approach} that apply state-of-the-art
visual question answering (VQA) models that make \technique applicable to a broad range
of datasets and trained models.

With the advent of high-quality text conditional generative models, like Flux~\cite{fluxmodel},
one might hope that simply
prompting with the text of the precondition will generate appropriate test inputs --
unfortunately this isn't the case.
Table~\ref{tab:sgsmsamples} shows three rows of randomly sampled images.
The top rows shows training samples selected to be consistent with the 
precondition.
The middle row shows
samples generated using  the best performing of a variety of base prompts that we explored, e.g., ``An image from a camera mounted at the top of a car's windshield'', and to which the text of the precondition was added.  
The first observation one can make is that the ``style'' of the prompted images does not match the training data, but there are other differences that can be observed as well, e.g., the prevalence of cars driving in the opposite direction of the ego
vehicle, and the shift vertical perspective of the camera.  If \technique were to return
inputs like this it is likely that developers of the LC under test would consider them
to be false positives, since they do not resemble images from the training dataset.

\begin{table}[t]
    \centering
    \begin{tabular}{c}
        \raisebox{5mm}{Training} 
        \includegraphics[width=0.215\linewidth, height=1.2cm]{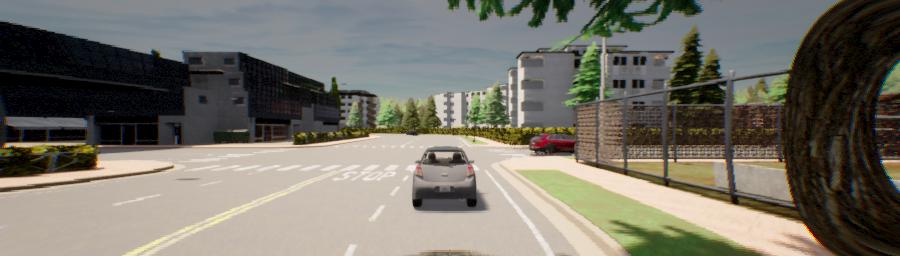}  
        \includegraphics[width=0.215\linewidth, height=1.2cm]{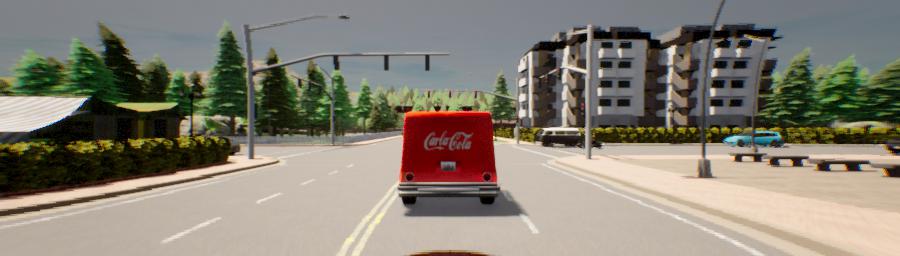}
        \includegraphics[width=0.215\linewidth, height=1.2cm]{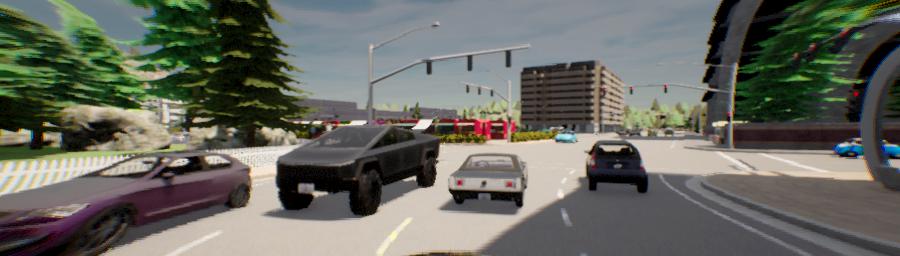}   
        \includegraphics[width=0.215\linewidth, height=1.2cm]{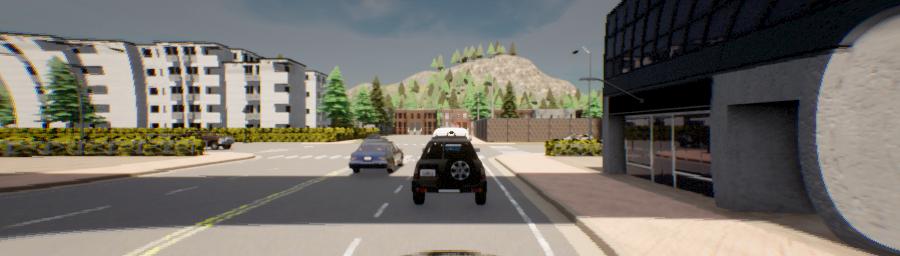} \\
        \raisebox{5mm}{Prompt} \hspace{1mm} \includegraphics[width=0.215\linewidth, height=1.2cm]{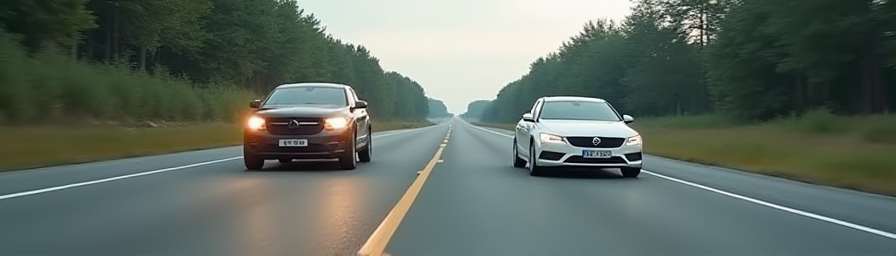}  \includegraphics[width=0.215\linewidth, height=1.2cm]{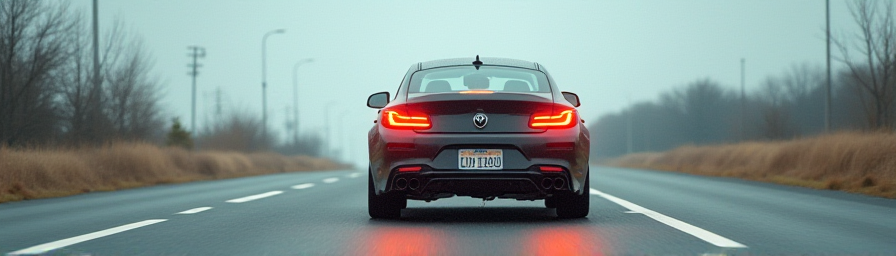} 
        \includegraphics[width=0.215\linewidth, height=1.2cm]{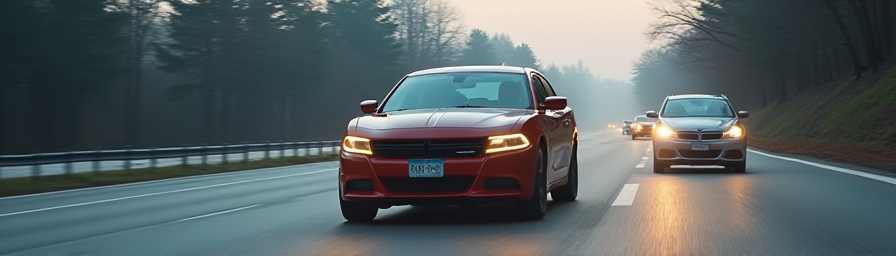} \includegraphics[width=0.215\linewidth, height=1.2cm]{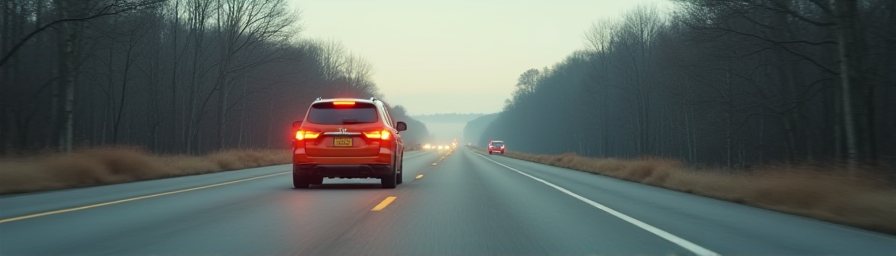}\\
        \raisebox{5mm}{LoRA} \hspace{4mm} \includegraphics[width=0.215\linewidth, height=1.2cm]{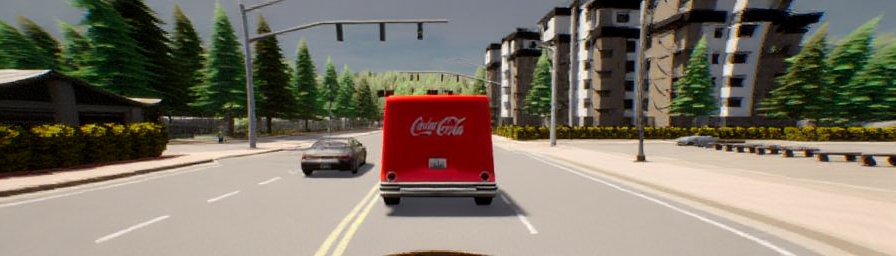}  \includegraphics[width=0.215\linewidth, height=1.2cm]{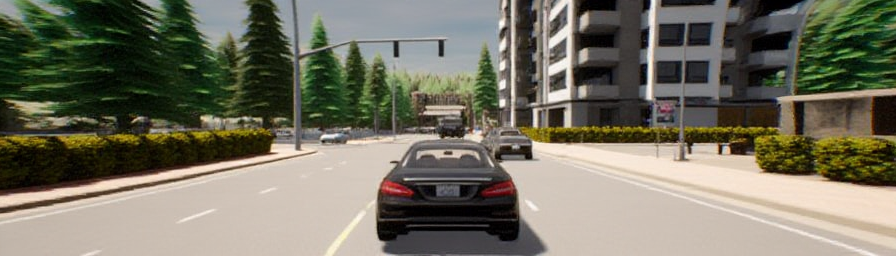} 
         \includegraphics[width=0.215\linewidth, height=1.2cm]{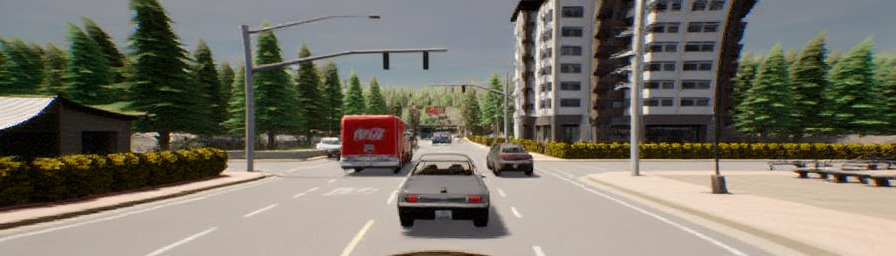}  \includegraphics[width=0.215\linewidth, height=1.2cm]{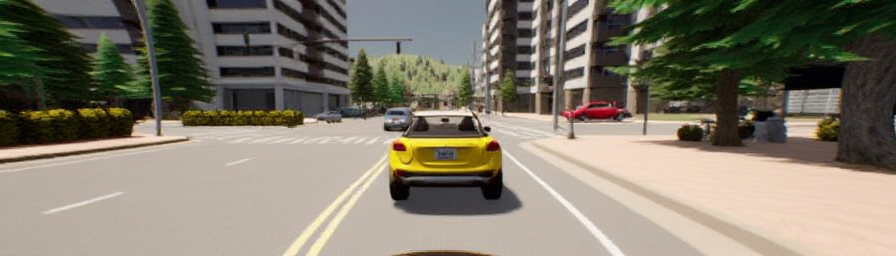}
    \end{tabular}
    \caption{Random samples from (top) SGSM training inputs filtered by the precondition \textbf{a vehicle is within 10 meters, in front, and in the same lane}, (middle) pre-trained Flux latent-diffusion model prompted with precondition, and (bottom) pre-trained Flux LoRA fine-tuned with precondition.}
    \label{tab:sgsmsamples}
\end{table}

Rather than simply prompting, we instead fine-tune a LoRA~\cite{hu2022lora} for Flux using the training data  subset
that is consistent with the precondition. 
The bottom row of Table~\ref{tab:sgsmsamples} shows random samples generated from such
a LoRA.
These images are \textbf{consistent} with the precondition, are \textbf{realistic} in comparison to training data, and are \textbf{diverse} in terms of the
structure of the road and the number and variety of vehicles on the road.
These properties of generated test inputs are important because:
(a) realistic inputs lie on the
data distribution on which the LC has been trained -- thereby avoiding misleading test results~\cite{dola2021distribution,berend2020cats};
(b) diverse inputs have the potential to provide broader coverage and
fault exposure of the LC~\cite{dola2022input,dola2024cit4dnn}; and
(c) the LC's output on precondition consistent inputs can be assumed to satisfy the postcondition of the requirement, which can be formulated as a test oracle and used to detect faults.
An evaluation in Section~\ref{sec:evaluation}, shows quantitative and qualitative
evidence that these observations appear to generalize across a
range of datasets and requirements.

When developers can specify necessary conditions for correct LC operation as
natural language functional requirements, \technique offers a novel and complementary
approach to fault detection.   The \technique approach can also be used in a more
general use case, where LC developers are interested in conducting focused exploratory
analysis of model behavior.  They simply describe the class of inputs of interest as
a precondition and state expected behavior for that class and let \technique 
generate and filter inputs to show them one's that have unexpected behavior.
In Section~\ref{sec:evaluation}, we conduct this kind of exploratory analysis
of model behavior for Imagenet trained classifiers where the definition of 
input classes is comprised of a combination of semantic features shared
by a group of animals, e.g., birds, and the expected behavior is that the
models output is one of the 59 different classes that define
species of birds that are used in Imagenet.   The potential value of this analysis is that \technique
will generate many
thousands of bird images can be generated and only present to the developer
a small sample on which their LC identifies them as something other than a bird.
The ability to cost-effectively focus developer attention on unexpected behavior
like this may help identifying limitations in training datasets or in model training.

The main contributions of this paper are:
(1) \technique -- the first test generation approach that formulates requirements of intended DNN behavior over a semantic feature space that is applicable to complex image models;
(2) a suite of strategies for generating SNL glossary term descriptions from complex image inputs; 
(3) an exploration of strategies for fine-tuning pretrained
text-conditional generative models to match requirement preconditions; 
(4) an evaluation on 25 requirements over 4 datasets
that shows \technique's potential to produce consistent, realistic, diverse test inputs; and
(5) demonstration that \technique can provide valuable feedback to developers in terms of fault-detection and exploratory analysis of unexpected behavior.

\section{Background and Related Work}
\label{sec:background}
We describe the background necessary to understand \technique and describe the most
closely related work and how it differs from the proposed approach.

\subsection{Neural Networks}
Developing an LC is a data-driven process that involves defining
a training dataset, $(x,y) \in D$, whose inputs, $D_x = \{ x : (x,y) \in D \}$, reflect a larger unspecified 
data distribution, $\X$, that is expected during LC deployment.
Training a neural network, $\DNN$, aims to closely approximate the unknown
\textit{target} function,
$f : \R^m \rightarrow \R^n$, exemplified by the data, i.e., $f(x) = y$ for 
training data $(x,y)$.
We consider an LC to be a black box and focus on its input-output
behavior, but we note that 
 training  seeks to generalize the learned approximation beyond
the training data to the unseen deployment distribution~\cite{goodfellow2016deep}, and DNN testing should
generalize to that distribution as well~\cite{berend2020cats,dola2021distribution}.

\subsection{Functional Requirements in Structured Natural Language}
\label{subsec:SNL}
Natural language is commonly used to express requirements, because it can
be understood by nearly all stakeholders~\cite{pohl2010requirements}.
The use of unstructured natural language phrasing has been shown, however,
to have negative impacts on the quality and utility of resulting requirements, e.g.,
vagueness, inconsistency, ambiguity, duplication, errors of omission, and a lack of testability~\cite{mavin2010big,fernandez2017naming}.
To address this, researchers have developed and evaluated
a variety of \textit{structured natural language} (SNL)
methods for expressing requirements~\cite{pohl2010requirements,mavin2009easy,grosser2023comparative},
and applied such approaches to domain-specific expression of 
functional requirements~\cite{uusitalo2011structured,veizaga2021systematically}.

We focus on 
functional requirements expressed using 
an \textit{if-then-shall} template of the form:
\begin{quoting}
\textit{If} \textbf{precondition}, \textit{then the LC shall} \textbf{postcondition}
\end{quoting}
\noindent where \textbf{postcondition} describes constraints on the LC output
that is computed for inputs that satisfy the constraints defined in the \textbf{precondition}.
We leave other templates to future work.

A functional requirement like this is typically partial in that 
it describes a \textit{necessary condition} for correct system
behavior.   For example, a precondition need not cover the entire
input domain and a postcondition may describe a set of allowable
outputs instead of a specific output.
It is a best-practice to define these conditions using combinations
of predefined domain-specific \textit{glossary} terms~\cite{lamsweerde2009requirements,pohl2010requirements}.
For functional requirements, glossary terms can be viewed
as defining atomic propositions over inputs and
preconditions as logical combinations of those propositions~\cite{lamsweerde2009requirements}. 
For the example in \S\ref{sec:intro}, the conjunction of propositions defining distance,
direction, and lane occupancy can be expressed in SNL as \textbf{a vehicle is within 10 meters, in front, and in the same lane}.

\subsection{Neural Network Requirements}
Researchers have identified the need for requirements engineering approaches to adapt to the characteristics of machine learning~\cite{seshia2018formal,rahimi2019toward,vogelsang2019requirements}.
Such requirements may include functional and non-functional requirements
as well as process-related requirements.

We focus on functional requirements in this work, but we recognize that
in machine learning it is assumed that
the target function cannot be precisely defined~\cite{goodfellow2016deep}.
Despite the inability to completely express the desired behavior
of an LC, researchers have understood that partial information
about the target function can be leveraged for validating trained LCs.
Towards this end, Seshia et al.~\cite{seshia2018formal} defined 
11 classes of formal requirements with two very broad classes that are applicable to LCs that
approximate functions: input-output robustness and input-output relation.

The  majority of the neural network testing and analysis
literature has focused on input-output robustness properties.
These include: domain-specific metamorphic properties used
for testing~\cite{deeptest, demir2024test, huang2024neuron}, metamorphic properties that capture
plausible variability in sensor inputs of autonomous 
driving systems~\cite{hu2022if,hu2022check}, and more general
approaches that validate local robustness~\cite{deepxplore, dlfuzz, lee2020effective, wang2022bet}.

Robustness alone is not enough to define necessary conditions for a correct LC, which is why
Seshia et al. defined the more general input-output relation class.
Formal frameworks for expressing general
input-output relations in the LC input space have been defined, e.g.,~\cite{shriver2021reducing}, but these have proven 
difficult to use for expressing high-level features of inputs.
For example, defining a semantic feature like \textbf{in the same lane}
requires encoding all of the myriad ways lanes may appear in
an image pixel-map, e.g., solid, dashed and double lane lines, 
the aging of paint on lane lines, the curvature
of the lane, and variation in surface reflectivity and lighting, etc..
This is why Seshia et al.~\cite{seshia2018formal} call out 
the need to lift requirements to a \textit{semantic feature space}
that is appropriate for the problem domain.

In this paper, 
rather than attempt to formalize preconditions in the sparse, high-dimensional, and uninterpretable input space of an LC, 
we leverage learned embeddings of glossary terms to localize regions
in the input space associated with domain-specific features.  In \S\ref{sec:approach} we show how this allows
formulation of requirements that  express both  input-output robustness
and  input-output relations in a semantic feature space for a range of datasets.

\subsection{Generative Models}
A generative model is trained on an input dataset, $D_x$ 
to produce unseen data from the broader distribution, $x \not\in D_x \wedge x \sim \X$.
A common strategy for training such models is to define 
an encoder, $\Encoder$, and decoder, $\Decoder$, and train them to
reconstruct inputs,
$\min_{x \in D_x} \lVert x - \Decoder(\Encoder(x)) \rVert$.
A variety of encoder-decoder approaches have been developed that define a low-dimensional
latent space that follows a
standard normal distribution, $\Normal$~\cite{kingmavae,rahimi2019toward,vahdat2020nvae,child2021very,rombach2022high}.
Sampling from this latent space, $z \sim \Normal$,
and running the decoder, $\Decoder(z)$, generates unseen data from $\X$
with high-probability.

One such class of models, \textit{latent diffusion models} (LDM), incorporates trainable
cross-attention layers that learn to condition the generation process
based on embeddings computed for a text prompt.
This strategy has established the state-of-the-art with the vast majority of
the Hugging Face leaderboard for the ``text to image'' task comprised of
instances of LDMs~\cite{huggingfaceleaderboard}.
A second advantage of LDMs is that they can be pre-trained as foundation-models for
the text to image task and then fine-tuned using a variety of strategies to
make them better suited to a domain-specific image generation task.
A particularly efficient form of fine-tuning uses \textit{low rank adaptation} (LoRA)
which freezes the pre-trained parameters  and adds trainable decomposition
matrices that are much smaller and therefore more efficient to train~\cite{hu2022lora}.
While \technique could use any LDM and fine-tuning strategy, in this
work we explore the use of the
Flux~\cite{fluxmodel} model fine-tuned using LoRA for precondition specific data, as it is among the best performing open source models~\cite{huggingfaceleaderboard}.

\subsection{Test Input Generation for Learned Components}
Like any software component, a trained LC must be tested to determine if it is
fit for deployment.   This involves selecting a set of test inputs and for each input
defining the expected LC behavior. The current test generation techniques use either pixel-level transformations or feature-level variations to generate test inputs~\cite{deeptest, deepxplore, dlfuzz, lee2020effective, wang2022bet, DeepHyperion-CS, dola2024cit4dnn}. The techniques including pixel-level transformations use image transformations such as brightness, blur, rotation, and translation to generate test inputs~\cite{deeptest, demir2024test, huang2024neuron}.
A wide-range of feature-level manipulation based test input generation techniques have been developed in the literature~\cite{deephyperion, DeepHyperion-CS, byun2020manifold,byun2021black,sinvad,dola2024cit4dnn}. 
Approaches such as DeepHyperion~\cite{deephyperion, DeepHyperion-CS} use manual-expertise to identify the interpretable features of the training dataset and manipulate the features to generate test inputs, whereas others leverage generative models~\cite{byun2020manifold,byun2021black,sinvad,dola2024cit4dnn}.
The most recent of these methods has been shown to be capable of generating inputs that are
realistic and diverse with respect to the training data~\cite{dola2022input}, but
a limitation of these approaches is that they have no way to target a precondition.

Unlike this prior work, \technique uses text-conditional generation to  produce test inputs for an LC that target regions of its input
domain that represent combinations of semantic features relevant
to stated requirements.  As we show
in \S\ref{sec:evaluation}, generated test inputs frequently
satisfy stated preconditions which means that the LC output for those
inputs can be checked against
the postconditions to detect faults or provide confidence that
an LC meets the stated requirements.

\section{Approach}
\label{sec:approach}

\begin{figure}
\pgfdeclarelayer{background}
  \pgfsetlayers{background,main}
  \scalebox{0.9}{
  \begin{tikzpicture}
    \node[name=dataset] {$D_x$};
    \node[name=requirements, above of=dataset, align=center, yshift=1mm, xshift=-8mm] {$(p_i,q_i) \in R$};
    \node[name=properties, below of=dataset, align=center, yshift=5mm, xshift=-11mm] {$(\pre{i}, \post{i})$};
    
    \node[name=gtgen, operator, right of=dataset, align=center, xshift=5mm] {Label\\Terms};
    \node[name=filter, operator, right of=gtgen, align=center, xshift=13mm] {Filter\\by Pre};
    \node[name=train, operator, right of=filter, align=center, xshift=10mm] {Fine\\Tune};
    \node[name=genmodel, above of=train, yshift=4mm] {$\Decoder(\theta_0)$};

    \node[name=sample, operator, right of=train, align=center, xshift=15mm] {Generate\\Tests};
    \node[name=approaches, below of=sample, yshift=-2mm] {$\Normal$};

    \node[name=pretests, right of=sample, align=center, yshift=0mm, xshift=17mm] {$T_{p_i}$};

    \node[name=model, below of=pretests, yshift=0mm, xshift=6mm] {$\forall t \in T_{p_i} : \post{i}(N(t))$};

    \draw[->, dashed] ($(requirements.south)+(-5mm,0)$) -- ($(properties.north)+(-3mm,0)$);
    \draw[->, dashed] ($(requirements.south)+(-1mm,0)$) -- ($(properties.north)+(3mm,0)$);
    \draw[->] (requirements.east) node[above, xshift=12mm] {$\{ p : (p,q) \in R \}$} -|  (gtgen.north);

    \draw[->] (dataset.east) -- (gtgen.west);
    \draw[->] (gtgen.east) -- node[above,xshift=-1mm] {$D^g_x$} (filter.west);

    \draw[->] ($(properties.south)+(-3mm,0)$) -- +(0,-2mm) -| (filter.south);
    \draw[->] (filter.east) node[above, xshift=5mm, align=center] {$D^{p_i}_x$} -- (train.west);

    \draw[->] (genmodel.south) -- (train.north);
    \draw[->] (train.east) -- node[above, name=tuned] {$\Decoder(\theta)$} (sample.west);

    \draw[->] ($(requirements.north)+(-5mm,0)$) --+(0,4mm) -| node[right, yshift=-2mm] {$p_i$} (sample.north);

    \draw[->] (approaches.north) -- node[left] {$z$} (sample.south);
    \draw[->] (sample.east) -- (pretests.west);
    \draw[->, dotted, thick] ($(properties.south)+(3mm,0)$) --+(0,-9mm) -| ($(model.south)+(2mm,0)$);
    \draw[->, dotted, thick] (pretests.south) -- ($(model.north)+(-6mm,0)$);

    \node[right of=pretests, xshift=2mm, yshift=-3mm, align=center] {\large Run Tests};

\begin{pgfonlayer}{background}
   \path[fill=blue!20] ($(requirements.north east)+(-4mm,3mm)$) rectangle ($(gtgen.south east)+(6mm,-2mm)$);

   \path[fill=gray!20] ($(filter.north west)+(-5.25mm,5mm)$) rectangle ($(train.south east)+(2mm,-8mm)$);
   
   \path[fill=green!20] ($(sample.south west)+(-10mm,-9mm)$) rectangle ($(sample.north east)+(9mm,4mm)$);
      
   \path[fill=red!20] ($(sample.north east)+(9mm,0mm)$) rectangle ($(model.south east)+(2mm,-2mm)$);
\end{pgfonlayer}   

\end{tikzpicture}
} 
\vspace{-4mm}
\caption{The phases of \technique: (1) glossary term labeling (blue), (2) training a requirements conditioned generative model (gray), (3) generating a precondition-specific test suite (green), and (4) running tests to check the postcondition oracle (red).}
\label{fig:approach}
\end{figure}
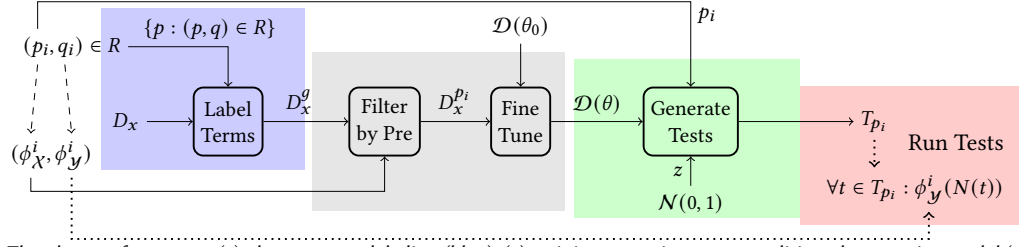

Figure~\ref{fig:approach} sketches the main elements of \technique.  It takes as input a set of
structured natural language (SNL)
statements, $R$, describing functional requirements for an LC, $N$,
defined over input domain $X$ and output domain $Y$.
It produces as output a set of test inputs, $T_{p_i}$, that is customized 
for the precondition, $p_i$, of a  requirement, $(p_i,q_i) \in R$.
The generated inputs, $T_{p_i}$, can be used to evaluate the behavior of the LC relative to a predicate, $\post{i}$, that encodes the postcondition, $q_i$.  If an input causes the
LC to violate the postcondition, then a fault, or unexpected behavior, has been detected.

We assume each requirement follows a template structure that allows for the identification
of the requirement precondition, $p$, and postcondition, $q$,
so we refer to requirements as a pair, $(p,q)$.
Moreover, we assume that requirement pre and postconditions are
expressed as the logical combination of glossary terms, $G$, that define semantic features as Boolean propositions over the input and output domain~\cite{arora2016automated}.

\technique operates in a series of four phases.
The first phase, shown in blue, labels
each element of the input dataset, $D_x$,
with a set of glossary terms that define the \textit{atomic} features referenced in system requirements, e.g., 
\textbf{a vehicle is within 10 meters}.
Glossary term definitions are domain specific and there are many possible strategies for performing this labeling process -- we describe two below.
This phase produces a glossary term labeled dataset, $D^g_x$.
The second phase, shown in gray, 
trains a text-conditional generative model, 
$\Decoder(\theta_0)$, on a dataset that is filtered using the glossary terms for 
training data, $D^g_x$, to evaluate the logical combination of terms defined by 
the precondition, $\pre{i}$.
The third phase, shown in green, leverages the fine-tuned decoder, $\Decoder(\theta)$.
A test input can be generated for a requirement precondition, $p_i$, by evaluating
$\Decoder(\theta)(p_i,z)$ where $z$ is sampled from the latent space of the generative model.
The final phase, shown in red, runs the generated tests and checks the postcondition oracle on model output,
$\forall t \in T_{p_i} : \post{i}(N(t))$.

The goal of \technique is to produce tests
that are (1) \textit{realistic} in comparison to the training dataset;
(2) \textit{consistent} with the requirement precondition; and
(3) as \textit{diverse} as $D_x$ subject to the constraints of $p_i$.
In the rest of this section, we describe how \technique is designed
to meet these goals and in \S~\ref{sec:evaluation} we evaluate
whether it meets those goals.

\subsection{Requirements}
\technique can work with a broad range of requirements descriptions.
Any logical combination of semantic feature descriptions can be used to
formulate a precondition, but to align with modern approaches to requirements
engineering we explore methods that use structured natural language.
In this paper, we explore SNL expression of two classes of requirements
that 
prior research~\cite{seshia2018formal} has identified as being useful:
semantic feature robustness requirements, and semantic feature functional requirements.

Both types of requirements are expressed using a domain-specific set of terms that is defined by LC developers.  We note that for systems
that already follow state-of-the-practice requirements engineering
approaches this set of terms has already been defined prior to 
the testing process.

Terms represent a semantic feature that may appear in LC inputs.
The current formulation of \technique supports Boolean terms.
For example, depending on the dataset one can express the presence of features like an \textbf{animal has feathers}, a \textbf{person has black hair}, or a \textbf{vehicle is in the same lane} in the input.

Features that are not naturally modeled as Boolean quantities such as the degree of lean to a digit or the distance that a vehicle lies from the ego vehicle are modeled as sets of disjoint Boolean features that can be combined logically.  For example, writing that \textbf{a vehicle is within 10 meters} is shorthand for \textbf{a vehicle is within 4 meters, or between 4 and 7 meters, or between 7 and 10 meters} -- the combination of three Boolean terms.

\ignore{
To simplify the expression of requirements we allow several shorthands.
For sets of features that are disjoint, when one is mentioned
positively the rest are implicitly negated.  For example, use of the term
\textbf{black hair} implicitly conjoins \textbf{not brown hair},
\textbf{not blond hair}, and \textbf{not bald}.
For features that are ordered, we allow expression of disjunctions
using ``within'', ``beyond'', ``less than'', or ``greater than''.
For example, \textbf{within 10 meters} represents \textbf{within 4 meters},
\textbf{between 4 and 7 meters, or} \textbf{between 7 and 10 meters}.
}

This disjunction of terms is implicit, but arbitrary
logical combinations of terms can be expressed in SNL using:
conjunction (disjunction) of terms expressed in comma-separated lists
ending  with ``and'' (``or'') -- for lists of length two the comma is dropped; and
negation expressed using ``no'', ``not'', or ``does not''.

We note that \technique can accommodate any standardized approach
to SNL expression of term combinations, but in this paper we use the
style described above.

\begin{table}
\footnotesize
\centering
\resizebox{\textwidth}{!}{
\begin{tabular}{c|c|c|p{0.55\textwidth}|p{0.25\textwidth}}
     & Id & Type:Src & Precondition & Postcondition \\ \toprule
\multirow{8}{*}{\rotatebox{90}{MNIST}}  & M1 & CFSR & The \textcolor{red}{digit is a 2} and \textcolor{orange}{has very low height} & label as 2\\ 
     & M2 & CFSR & The \textcolor{red}{digit is a 3} and \textcolor{orange}{is very thick} & label as 3\\  
     & M3 & CFSR & The \textcolor{red}{digit is a 7} and \textcolor{orange}{is very thick} & label as 7\\ 
     & M4 & CFSR & The \textcolor{red}{digit is a 9} and \textcolor{orange}{is very left leaning} & label as 9\\ 
     & M5 & CFSR & The \textcolor{red}{digit is a 6} and \textcolor{orange}{is very right leaning} & label as 6\\ 
     & M6 & CFSR & The \textcolor{red}{digit is a 0} and \textcolor{orange}{has very low height} & label as 0\\ 
     & M7 & CFSR & The \textcolor{red}{digit is an 8} and \textcolor{orange}{is very thin} or \textcolor{blue}{very thick} & label as 8\\ \midrule
\multirow{8}{*}{\rotatebox{90}{Celeba-HQ}} & C1 & CFSR & The \textcolor{red}{person is wearing eyeglasses} and  \textcolor{orange}{has black hair} & label as eyeglasses\\ 
     & C2 & CFSR & The \textcolor{red}{person is wearing eyeglasses} and \textcolor{orange}{has brown hair} &  label as eyeglasses\\ 
     & C3 & CFSR & The \textcolor{red}{person is wearing eyeglasses} and \textcolor{orange}{has a mustache} & label as eyeglasses\\ 
     & C4 & CFSR & The \textcolor{red}{person is wearing eyeglasses} and \textcolor{orange}{has wavy hair} & label as eyeglasses\\ 
     & C5 & CFSR & The \textcolor{red}{person is wearing eyeglasses} and \textcolor{orange}{is bald} & label as eyeglasses\\ 
     & C6 & CFSR & The \textcolor{red}{person is wearing eyeglasses} and \textcolor{orange}{ a hat} & label as eyeglasses\\  
     & C7 & CFSR & The \textcolor{red}{person is wearing eyeglasses} and \textcolor{orange}{ has a 5 o'clock shadow} or \textcolor{blue}{goatee} or \textcolor{magenta}{mustache} or \textcolor{green}{beard} or \textcolor{purple}{sideburns} & label as eyeglasses\\ \midrule
\multirow{11}{*}{\rotatebox{90}{SGSM}}  & S1 &  SFFR:\S46.2-816 & A \textcolor{red}{vehicle is within 10 meters}, \textcolor{orange}{in front}, and \textcolor{blue}{in the same lane}	& not accelerate\\ 
     & S2 & SFFR:\S46.2-833 & The \textcolor{red}{ego lane is controlled by a red} or \textcolor{orange}{yellow light}	& decelerate\\ 
     & S3 & SFFR:\S46.2-888 & The \textcolor{red}{ego lane is controlled by a green light},  and no \textcolor{orange}{vehicle is in front}, \textcolor{blue}{in the same lane}, and \textcolor{violet}{within 10 meters}	& accelerate\\ 
     & S4 & SFFR:\S46.2-802 & The \textcolor{red}{ego is in the rightmost lane} and not \textcolor{orange}{in an intersection}	& not steer to the right\\ 
     & S5 & SFFR:\S46.2-802 & The \textcolor{red}{ego is in the leftmost lane} and not \textcolor{orange}{in a intersection}	& not steer to the left\\ 
     & S6 & SFFR:\S46.2-842 & A \textcolor{red}{vehicle is in the lane to the left} and \textcolor{orange}{within 7 meters}& not steer to the left\\ 
     & S7 & SFFR:\S46.2-842 & A \textcolor{red}{vehicle is in the lane to the right} and \textcolor{orange}{within 7 meters} & not steer to the right\\ \midrule
 \multirow{4}{*}{\rotatebox{90}{ImageNet}}  & I1 & SFFR:\cite{miller2018zoology} & The \textcolor{red}{single real animal} \textcolor{orange}{has feathers}, \textcolor{blue}{wings}, \textcolor{purple}{a beak}, and \textcolor{green}{two legs}	& label as a hyponym of bird\\
     & I2 & SFFR:\cite{miller2018zoology} & The \textcolor{red}{single real animal} \textcolor{orange}{has fur or hair}, \textcolor{blue}{hooves}, and \textcolor{purple}{four legs}	& label as a hyponym of ungulate\\
     & I3 & SFFR:\cite{miller2018zoology} & The \textcolor{red}{single real animal} \textcolor{orange}{has an exoskeleton}, \textcolor{blue}{antennae}, and \textcolor{purple}{six legs}	& label as a hyponym of insect\\
     & I4 & SFFR:\cite{miller2018zoology} & The \textcolor{red}{single animal} \textcolor{orange}{has} no \textcolor{orange}{limbs} and no \textcolor{blue}{ears}& label as a hyponym of snake\\
     \bottomrule
\end{tabular}}
\caption{Requirement preconditions and postconditions for four datasets spanning two types of properties: conditional semantic feature robustness (CSFR) and
semantic feature functional requirements (SFFR).  Distinct glossary term phrases are highlighted with colors within each precondition.  The remaining text and punctuation, shown in black, defines the logical combinations of glossary terms.}
\label{tab:requirements}
\end{table}

\subsubsection{Feature-based Robustness}
Assessing LC robustness involves determining that output is consistent with limited perturbation of an input.
Conventional approaches primarily focus on low-level input perturbations such as changes in pixel values, lighting, or spatial transformation to assess the robustness, e.g.,\cite{deeptest, demir2024test, huang2024neuron}.
These techniques fail to consider semantic-feature variation.

Ideally, the LC's output should remain consistent when the input is varied with  respect to semantic features that are irrelevant to the decision objective. 
For example, the satisfaction of the requirement illustrated in the
LoRA rows of \autoref{tab:sgsmsamples} should be independent of the make
of the vehicle, e.g., whether it is a truck or a car, or its color, e.g.,
whether it is black, silver, or grey.  These semantic features
are \textit{independent} of the decision problem, i.e., whether to accelerate
or not.

More formally, the output of an LC is robust with respect to a set of semantic features, $I$, if its output is invariant to perturbations those features.   
\begin{definition}[Global Semantic Feature Robustness (GSFR)]
Given an LC $N : X \mapsto Y$, a set of features $I$, and an operator,
$\oplus : X \times I \mapsto X$, that can perturb inputs with respect to 
features we say the LC is \textbf{globally semantic feature robust} if:
\begin{equation*}
    \forall x \in X: \forall i \in I :  N(x) = N(x \oplus i)
\end{equation*}
\end{definition}
Many variants of this definition can be formulated.  For example, one
can restrict the inputs to satisfy a specific constraint which allows
the definition of $I$ to be specialized to that class of inputs.
\begin{definition}[Conditional Semantic Feature Robustness (CSFR)]
Given an LC $N : X \mapsto Y$, a predicate $\phi \subseteq 2^X$, a set of features $I$, and an operator,
$\oplus : X \times I \mapsto X$, that can perturb inputs with respect to 
features we say the LC is \textbf{conditional semantic feature robust} if:
\begin{equation*}
    \forall x \in X: \phi(x) \implies \forall i \in I :  N(x) = N(x \oplus i)
\end{equation*}
\end{definition}

\technique allows both $\phi$ and $I$ to be expressed using SNL phrasing.
To illustrate consider an MNIST digit classifier with
\[ I = \{\text{``very left leaning''}, \text{``left leaning''}, \text{``right leaning''}, \text{``very right leaning''}\}\]
describing digits that are not upright; $I$ can be interpreted as a disjunction
of semantic features.    The individual elements of $I$ define equivalence
class of ``lean angle'' as described below so each describes variability within the named class.  A GSFR requirement would express that the digit classification task is invariant to the direction or degree of lean of the digit.   

\technique
can be applied to generate test inputs to validate a GSFR property, but in this 
paper we focus on the more challenging CSFR requirement because they demonstrate the
ability of \technique to restrict semantic feature variation to subsets of the
LC input space.
To illustrate, consider requirement M1
 in Table~\ref{tab:requirements}.
It restricts inputs using $\phi = \text{``digit is a 2''}$ and uses a singleton
semantic feature set $I = \{ \text{``very low height''} \}$.
Note that $\phi$ defines a set of possible inputs, all those consistent
with $\phi$, and that $I$ also defines a set of semantic feature variation
that lie within the set of heights defined as being ``very low''.

Since we selected a restriction on digits for $\phi$ the robustness 
property can be expressed using a postcondition that simply expects the
corresponding digit class.  In essence, the correspondence between
$\phi$ and the postcondition specifies the value computed by $N(x)$
in the GSFR and CSFR definitions above.

Requirements M1-M7 select different $\phi$ that correspond to different MNIST
digits and with this restriction variation in independent features means that the
output class should be the digit described by the constraint which we depict
here as the postcondition.  We vary $I$ here to capture different directions
and degrees of slant and thickness in digits; including disjunctive constraints which can be interpreted as $I$ being a set of size greater than one.

Requirements C1-C7 use the same $\phi$ to restrict the inputs to those where
a person is wearing eyeglasses and with this restriction the output should
indicate the presence of eyeglasses.   We vary $I$ here to capture variation in the appearance of a person's head, e.g., hair color, texture, the lack of hair, or the presence of different types of facial hair.

The ability to use semantic features, for both $\phi$ and $I$, allows for 
a  broad class of inputs to be generated by \technique
as can be seen from images generated from the C4 precondition in Fig.~\ref{fig:rq1_celebahq_samples}.  This 
variability would be impossible to achieve with pixel-based robustness testing.

Our selection of robustness properties is not meant to be exhaustive or
to necessarily reflect what would be of interest to MNIST or CelebA-HQ LC developers, rather our goal is to demonstrate how concise SNL expressions
can be used to flexibly focus test generation to achieve new forms of
robustness testing.

\subsubsection{Semantic Feature Functional Requirements (SFFR)}
A general class of requirements identified by prior research~\cite{seshia2018formal} allows for the expression of 
functional LC behavior, i.e., acceptable input-output relations.

Unlike robustness requirements, functional requirements can levy a richer class of constraints on 
the output space of the LC.
For categorical LCs glossary terms 
express whether a label is in a set of possible labels or not, e.g.,
\textbf{label as 6} or \textbf{label is robin, ..., or eagle}.
One can define domain-specific notational shorthands to allow for 
more concise requirements statements, for
example, we write \textbf{label is hyponym of bird} for the 59-way disjunction of classes that lie
below ``bird'' in the WordNet taxonomy.
For regression LCs glossary terms define constraints on the output values, e.g.,
\textbf{accelerate} equates to $N(\cdot).accel > 0$ and \textbf{not steer to the right}
to $N(\cdot).steer \ge 0$.

Given an \textit{if-then-shall} SNL requirement statements 
we can extract the logical combination of feature terms
present in the precondition, $\pre{}$, and the logical
combination of feature terms present in the postcondition,
$\post{}$.  These are the building blocks for defining a functional requirement.
\begin{definition}[Semantic Feature Functional Requirement (SFFR)]
Given an LC $N : X \mapsto Y$, a precondition $\pre{} \subseteq 2^X$, 
and a postcondition $\post{} \subseteq 2^Y$, we say the LC meets the \textbf{semantic feature functional requirement} if:
\begin{equation*}
    \forall x \in X: \pre{}(x) \implies \post{}(N(x))
\end{equation*}
\end{definition}
\technique aims to generate inputs that satisfy $\pre{}$ upon which
the output of the LC can be checked with $\post{}$ and thereby validate
the requirement.

Requirements S1-S7 in \autoref{tab:requirements} depict functional 
requirements for the input-output
relation of an autonomous driving LC trained on the SGSM dataset.  
Each requirement encodes
a necessary condition for a section of the Virginia driving code, e.g., 
\S46.2-842.   Here the precondition terms express semantic features of the lane in which the ego vehicle is driving and distance, direction, and lane occupancy information of other vehicles relative to the ego vehicle.   The postcondition terms express semantic features of the acceleration and steering angle outputs of the LC.

Whereas the SGSM requirements describe necessary conditions whose violation can reasonably thought of as faults, requirements I1-I4 express expected behavior that allows developers of ImageNet trained LCs to explore questions
about model behavior.
For example, 
 I1 expresses an input-output relation defining the expectation that if an ImageNet input has a set of features that are characteristic of birds then an LC will return one of the WordNet~\cite{fellbaum2010wordnet} terms for bird. This is expressed in SNL as:
\begin{quoting}
\textit{If} \textbf{the single real animal has feathers, wings, a beak, and 2 legs}\textit{, then the LC shall} \textbf{label as a hyponym of bird.}
\end{quoting}
\noindent Here the precondition expresses a conjunction of glossary
terms describing morphological features, e.g., \textbf{has feathers}, present in the image that in combination discriminate zoological taxa~\cite{miller2018zoology}.
The first term, \textbf{single real animal}, aims to preclude paintings or other representations of birds, and diverse groupings of birds from matching the precondition.
The postcondition restricts the set of allowable labels that the
LC may produce to those below ``bird'' in the WordNet taxonomy~\cite{fellbaum2010wordnet}.
The premise of such a requirement is that a correct model should at least be
able to classify a bird as \textbf{some kind of bird}, and if it cannot then a developer would want to explore why that is the case.

\subsection{Glossary Term Labeling and Annotation}
\label{subsec:glossary_term_labeling}
Data labeling is an essential problem in supervised learning and we view glossary
term labeling as an instance of that problem.
While manual labeling is considered to be a robust representation of ground truth, the
cost of labeling has led to the development of a number of automated labeling approaches, e.g.,
~\cite{ratner2020snorkel,das2020goggles,zhou2024openannotate3d,zhang2024recognize}.
\technique is agnostic to the particular method used to perform glossary term labeling, but
like other labeling problems the more accurate the labeling the better.
In this section, we outline two general strategies for labeling data with
glossary terms: (1) post-processing the output of existing analytic
methods for a dataset, and (2) prompting visual question answering (VQA) models~\cite{antol2015vqa}.

Using the first of these approaches depends on the existence
of an analytic method for a dataset.
We illustrate this approach by reusing an existing analytic method
for the MNIST dataset and by developing a new analytic method
for the SGSM autonomous driving dataset~\cite{woodlief2024s3c,sgsm}.
When no such analytic method exists, we explore the the use
of broadly applicable ML-based auto-labeling strategies and describe one
such method below.

Given a glossary, $G$, and an input domain, $X$, 
a \textit{glossary term labeler}, is a function
$gt : X \mapsto 2^{E \times 2^G}$, where $E$
is the set of entities present in the input.
This definition supports
input domains that allow multiple entities to be present in
a single input, e.g., multiple vehicles; if a single entity
is present then the codomain of $gt$ reduces to $2^G$.
\cite{rbt4dnn} provides complete details of the labeling process, along with the corresponding code implementation.


For MNIST, we reuse an analysis, $morpho$,
that computes a set of morphological measures, $M$, of digits such as:
thickness, slant, and height~\cite{castro2019morpho}.
Slant is defined based on the angle of a parallelogram bounding the
digit with vertical defining a reference of 0.
The remaining characteristics are all defined as distances
based on the parallelogram.
We partition the value ranges for these measures,
$part : X \times M \mapsto g$, to
define terms like \textbf{very thick} and \textbf{very right leaning}
for this dataset, $gt(x) = \{ g : part(morpho(x),m) \wedge m \in M \}$,
since there is a single entity in the input -- the digit.

\ignore{
\begin{wrapfigure}{r}{0.53\linewidth} 
\footnotesize
    \centering
    \resizebox{.5\textwidth}{!}{
    \begin{tabular}{c|c|c|l}
         Attr. & Term & Values & SNL Text  \\ \toprule
         \multirow{5}{*}{\rotatebox{90}{Thickness}} & VThin & $[0,1.5)$ & digit is very thin \\
         & Thin & $[1.5,3.5)$ & digit is thin\\
         & SThick & $[3.5,4)$ & digit is slightly thick\\
         & Thick & $[4,7)$ & digit is thick \\
         & VThick & $[7,\infty)$ & digit is very thick\\
         \midrule
         \multirow{4}{*}{\rotatebox{90}{Slant}} & VRight & $(-\infty,-0.4]$ & digit is very left leaning \\
         & Left & $(-0.4,-0.1)$ &  digit is left leaning \\
         & Upright & $[-0.1,0.1]$ &  digit is upright \\
         & Right & $(0.1,0.4)$ &  digit is right leaning\\
         & VRight & $[0.4,\infty)$ &  digit is very right leaning\\
         \midrule
         \multirow{4}{*}{\rotatebox{90}{Height}} & VLow & $[0,14)$ &  digit has very low height\\
         & Low & $[14,17)$ &  digit has low height\\
         & High & $[17,20)$ &  digit has high height\\
         & VHigh & $[20,\infty)$ &  digit has very high height\\
         \bottomrule
    \end{tabular}}
    \caption{Glossary \textbf{term}s for MNIST digits for different ranges of \textbf{values} for different Morphometric \textbf{attr}ibutes with associated \textbf{SNL text} phrasing.}
    \label{tab:mnist_glossary}
\end{wrapfigure}
}

We defined a glossary term labeler for a more complex input domain
by building on the \textit{spatial semantic scene coverage} (S3C) framework
which defines an abstraction of pixel-based image data that can be mined to define a requirements glossary~\cite{woodlief2024s3c}.
More specifically, this abstraction defines a \textit{scene graph} whose vertices, $V$, 
represent instances of entities in the image that are relevant to requirements, e.g., ``car'', ``lane'', or ``traffic signal'',
and whose edges represent spatial relationships, $R$, among entities, e.g., ``in'', ``left of'', or ``within 4 to 7 meters''.
The graph can be encoded as set of triples, $(v,r,v')$ where $v,v' \in V$ and $r \in R$.

In scene graphs there is a special vertex representing the ``ego'' vehicle upon which the camera is mounted and all glossary terms are expressed relative
to that vehicle.
We define a function, $walk : V \times 2^P$, that
computes all acyclic paths, $P$, from the ego to a given vertex in the graph.
We define a function, $gp : P \times G$, which computes a glossary
term from a path.
For example, for the sequence of triples:
$(ego,in,lane1)$, $(lane2,\textit{leftof},lane1)$,
$(car17,in,lane2)$, 
the function would compute that $car17$ \textbf{is in the lane to the left}.
Reusing an existing scene graph generator, $sg$, 
allows us to define a glossary term labeler as:
$gt(x) = \{ (v,gp(p)) : (\_,\_,v) \in sg(x) \wedge p \in walk(v) \}$,
where the $\_$ expresses wildcard matching.

\ignore{
In general there will be domain-specific implicit constraints that impact the meaning of glossary terms.  In our evaluation,
we consider three datasets: CelebA-HQ, MNIST, and SGSM, which comprise inputs from very different domains.
In CelebA-HQ, which is comprised
of close up headshots of people, it is implicit that
the glossary terms
all modify the sole \textit{entity}, in this case a person, depicted in the image -- there is never a second entity in the image being described.
Similarly, for MNIST it is implicit that the input
represents a single entity, a digit, that
all glossary terms modify.
The SGSM dataset has richer constraints.
It is implicit that SGSM data are images from a forward
facing camera mounted the sole ego vehicle -- a sliver of the
ego hood
appears at the bottom of the image in the middle as shown
on the left of Figure~\ref{fig:image-sg-terms}.
It is also implicit that glossary terms that do not 
directly mention the ego vehicle, are relative to the ego
vehicle, e.g., ``A car is in front'' means ``A car is in front of the ego vehicle''.
In addition, there may be multiple instances
of entities in SGSM images, e.g., lanes, vehicles, traffic lights,
and multiple glossary terms may modify instances of those entities.
For such datasets, term labeling must reflect such entity-term associations, e.g., as shown at the bottom of Figure~\ref{fig:image-sg-terms} there are two cars in the image each with a different set of modifying glossary terms.
}

A third, more general, approach uses the textual representations of glossary
terms to form prompts for a VQA model.   To use a VQA we formulate a prompt of the form
``Does the object have \textbf{glossary term}? Answer only yes or no.'' for each glossary term, e.g.,
feathers, eyeglasses; we use slight variations of the prompt to make it more fluent.
There are a range of different pre-trained VQA models available and 
we found that the Mini-CPM~\cite{yao2024minicpm}
model was quite accurate in computing glossary term labels for
CelebA-HQ relative
to human annotations and for the ImageNet requirements we considered.
With a function mapping glossary terms to prompts, $prompt : G \mapsto String$,
this family of glossary term labelers is defined as:
$gt(x) = \{ g : g \in G \wedge vqa(prompt(g),x) = yes \}$.

\subsection{Training for \technique}
Given the glossary term labeled dataset, $D^g_x$, there are
a variety of strategies one might take to train a text-conditional
generative model.  
The architecture of the generative model and the training objectives
used to train it can influence
the quality of generated data, but we do not consider those
to be choices that are specific to \technique.
The training data, however, is specific to \technique.

\ignore{
Computing SNL text annotated data, $D^a_x$, provides a training
dataset pairing images and text that can be used to train from
scratch a generative model architecture or fine-tune a pretrained generative model.   
In preliminary work, we explored both strategies, but in Section~\ref{sec:evaluation} we evaluate
implementations that fine-tune large, e.g., 12 billion parameter, models that are pre-trained on
large and diverse datasets and trained to produce high-resolution and high-quality images.
Such models are readily available, e.g.,~\cite{huggingfaceleaderboard}, and
represent the current state-of-the-art.
}

Fine-tuning models can be achieved using a variety of strategies,
but it can be expensive to do so for very large models.
While such training can yield excellent results, we chose to
explore the use of low-rank adaptation (LoRA), which introduces a
small set of parameters that are trained during fine-tuning
and whose results are combined with the output of the 
pre-trained model to best match the fine-tuning data~\cite{hu2022lora}.
Many state-of-the-art models now come with pre-defined LoRA that are
designed to optimize the quality of generated images and 
training time.
In this work, we fine-tune using LoRA and leave the exploration of
cost-benefit tradeoffs for alternative training approaches to future work.

As depicted in Figure~\ref{fig:approach}, \technique uses a filtering approach to define image-text pairs for fine-tuning.
More formally, we define 
$D^{p_i}_x = \{ (x,p_i) : (x, gt(x)) \in D^g_x \wedge \pre{i}(gt(x)) \}$, which applies the precondition
to the glossary terms for an input; by construction
glossary terms directly map to valuations of atomic propositions in the precondition.
We note some important differences between this strategy and the more straightforward strategy of annotating
the entire training dataset.

First, when a precondition, $p_i$, describes a rare set of inputs
then $\lvert D^{p_i}_x \rvert \ll \lvert D_x \rvert$ and it is essential to fine-tune with a LoRA since they generally have lower-data requirements and fewer parameters to train.
Second, $D^{p_i}_x$ uses the precondition SNL as a text annotation which directly relates all of the training images to the precondition on which it will be ultimately prompted.
Third, fine-tuning \textit{per precondition} models, i.e., using
$D^{p_i}_x$, focuses fine-tuning on a single precondition which offers the potential for greater precondition consistency.  
This does, however, mean that a LoRA per requirement must be trained to instantiate
\technique for a given LC.
In \S~\ref{sec:evaluation}, we report on the relative performance of several different training strategies.

\section{Evaluation}
\label{sec:evaluation}

In this section, we describe the design and findings of our evaluation of \technique, focusing on two key aspects: the quality of the generated test suites and their applicability in assessing model robustness and behavior. To guide our evaluation, we define the following research questions:\\

\noindent\textit{Quality of Generated Test Inputs:}
This category examines the degree to which \technique-generated images holds the properties- consistency with requirements, realism and diversity- essential properties that are expected from a reliable test suit.
\begin{itemize}
    \item \nameref{sssec:rq1}
    \item \nameref{sssec:rq2}
    \item \nameref{sssec:rq3}
\end{itemize}

\noindent\textit{Applicability of Generated Tests:}
This category focuses on the practical use of \technique-generated test images in revealing faults in models and interpreting their behavior.
\begin{itemize}
    \item \nameref{sssec:rq4}
    \item \nameref{sssec:rq5}
    \item \nameref{sssec:rq6}
\end{itemize}


\subsection{Evaluation Design}
\label{subsec:design}
Our evaluation spans four datasets,  with 4-7 requirements per dataset, and three approaches
to fine-tuning a pre-trained generative model.  We use standard metrics to assess
how realistic images are for RQ2 and fault-detection effectiveness
for RQ4-RQ6.   The metrics for RQ1 and RQ3 
are based on the glossary terms that describe generated images and to compute those terms
we train sets of binary classifiers.
We describe the considerations leading to our experimental design below.

\subsubsection{Dataset Selection}
This work focuses primarily on the LCs that accept image inputs and compute categorical or regression outputs.
To instantiate \technique on a dataset we must be able to perform term labeling using a technique from \S\ref{subsec:glossary_term_labeling}.
Based on these requirements, we selected four datasets: MNIST~\cite{deng2012mnist}, CelebA-HQ~\cite{karras2018progressive}, SGSM~\cite{toledo2024specifying}, and ImageNet~\cite{deng2009imagenet}, that vary in complexity and domain.

CelebA-HQ is a high-resolution subsample of the CelebA ``headshot'' dataset\cite{liu2015deep}.
CelebA-HQ has human defined labels~\cite{CelebAMask-HQ} for 40 different features, e.g., hair color, gender, that we use as
glossary terms.  We apply VQA-based term labeling
using the MiniCPM-o-2\_6~\cite{MiniCPM2025} model, which ranks in the top-2
on Hugging Face as of January 2025.  With both generated glossary terms and human labels, we can directly
compare the performance of \technique with these two label sources.

We use a version of the MNIST  digit dataset that is upscaled to $64^2$ pixels to be compatible with the LDM used for image generation.
We leverage prior research that computes
morphometric measurements of digits to define features
like digit thickness, slant, and height~\cite{castro2019morpho}.
We partitioned those value ranges
to define glossary terms for formulating requirements over this dataset.

We built on the SGSM autonomous driving dataset~\cite{sgsm} which consists of 10885 900x256 pixel images from
a forward facing camera on a simulated ego vehicle in Town05 of the CARLA Autonomous Driving Leaderboard~\cite{carlaleaderboard}. 
The SGSM infrastructure defines scene graph abstractions for these images~\cite{woodlief2024s3c,toledo2024specifying}, e.g.,
Figure~\ref{fig:image-sg-terms}.  
We leverage the natural language phrasing of logical specifications used in this work~\cite{toledo2024specifying}
to define glossary terms.
To generate those terms, we use the technique from \S\ref{subsec:glossary_term_labeling} that performs a depth-first traversal, rooted
at the ego vehicle, to produce glossary terms describing each entity, e.g., vehicle, signal, lane, in the scene.

ImageNet~\cite{deng2009imagenet} is a large scale visual database with millions of images spanning a thousand categories.
Each category label is an element of WordNet~\cite{Miller1995}- a lexical database that organizes words into a taxonomy
based on semantic relationships.
We used standard morphological features of animals~\cite{miller2018zoology}, e.g., feathers, wings, hooves, or antennae,
that discriminate levels in the zoological taxa, e.g., birds, insects, and used those features as glossary terms.
We apply VQA-based term labeling
using the MiniCPM-o-2\_6~\cite{MiniCPM2025} model to label images with these terms.

\subsubsection{Choice of Requirements}
For MNIST and CelebA-HQ, we use the glossary terms to formulate feature-based robustness requirements.  
Robustness requires that we select inputs that will yield a known prediction.  To achieve
this, we select one feature that corresponds to the output prediction and pair it with at least one other feature to
form a precondition.  The postcondition asserts that the expected prediction is made by the model.
We could chose any pair of features, but here we explored pairs
that occur rarely in the dataset, e.g., below 1\% of the time.
From those we selected 7 at random for each dataset and we label them  M1-M7 and C1-C7 in Table~\ref{tab:requirements}.

For SGSM we formulated feature-based relational requirements in SNL based on properties from Table 1 in~\cite{toledo2024specifying}.
These properties define necessary conditions for safe driving under the Virginia Driving Code~\cite{CodeOfVirginia}.
The preconditions describe combinations of features within the vehicles field of view and the postconditions
define constraints on regression outputs of the LC, e.g., 
``accelerate'' by $N(\cdot).accel > 0$.
Table~\ref{tab:requirements} lists 7 such requirements, S1-S7, and associates them with the sections
of the Virginia Driving code from which they were derived, e.g., \S46.2-816.

For ImageNet we formulated the 4 requirements, I1-I4, in Table~\ref{tab:requirements} that use the zoological 
features described above.  We selected intermediate nodes in the taxonomic tree that corresponded to animals, e.g., ``bird'', and defined preconditions as combinations of morphological features defined in a Zoology textbook~\cite{miller2018zoology} for that animal.
The taxonomic relationships among words, e.g., that ``bird'' generalizes ``robin'', allows us to express necessary conditions for correct classification results; ``robin'' is a hyponym of ``bird''.
To do this we collect the leaves of the taxonomy rooted at a term and check membership in that set.
For example, let $bird$ denote the leaves of the taxonomy rooted at the word ``bird'', then a postcondition
``label as a hyponym of bird'' is checked as $N(x) \in bird$. 

\subsubsection{LC Selection}\label{ssec:lc_selection}
RQ4–RQ6 examine \technique’s effectiveness in assessing requirement-specific behavior, detecting faults, and revealing LC decision behavior.
We select LC's for each dataset that aim to be of high-quality relative
to the state-of-the-practice, e.g., they have high-test accuracy, use rich
architectures, etc.

For SGSM we trained an LC by extending an existing autonomous driving
model comprised of a ResNet34 architecture pretrained on ImageNet
data that was fine-tuned on data from the CARLA simulator~\cite{woodlief2024s3c}.   We note that the top-3
performing autonomous driving models on the CARLA leaderboard all use
a similar ResNet34 architecture.
The model we started with could
only predict the steering angle and we added an acceleration
prediction head, which allowed us to take advantage of the steering
and acceleration data generated by the CARLA simulations we
ran to produce image data -- as described above.
This model was trained to have a mean squared error (MSE) loss 
of 
0.010 (0.003 std) for steering angle and 
0.331 (0.067 std) for acceleration; MSE is a standard measure used in training regression models.

For MNIST rather than use one of the classic models for digit identification,
we chose to use a vision transformer model fine-tuned on the MNIST dataset from Hugging Face~\cite{hfmnist, wu2020visual}. This model has a test accuracy of 99.49\% which would place it in the top-30 of the MNIST leaderboard~\cite{mnistleaderboard}.

For the non-standard CelebA-HQ eyeglasses binary classification problem, we used a pre-trained vision transformer from Hugging Face~\cite{wu2020visual, vit-celeba} and trained it over the CelebA-HQ dataset, achieving an accuracy of 90.34\%.

For ImageNet, we selected three distinct ImageNet architectures: VOLO-D5~\cite{yuan2022volo} (V-D5) (87.1\%), CAformer-M36~\cite{yu2022metaformer_baselines} (C-m36) (86.2\%), and EfficientNet-B8~\cite{Xie2019AdversarialEI} (E-B8) (85.4\%). All of these models achieve over 85\% accuracy, which is within 6 percentage points of the top-performing model, CoCa (91\%), according to ImageNet benchmark~\cite{PapersWithCode2025}. 

\subsubsection{Fine-tuning Approaches}
For these experiments, we considered three approaches to fine-tuning
a pre-trained latent diffusion model.
We utilized the pretrained FLUX.1-dev~\cite{huggingface2024} model as our 
base model because it was the best performing open source model on
Hugging Face's text to image generation leaderboard~\cite{huggingfaceleaderboard} at the time we conducted the experiments.
Flux is a 12-billion-parameter rectified flow transformer capable of producing high-quality images from text descriptions.
To mitigate the cost of fine-tuning, we used a low-rank adaptation (LoRA)~\cite{hu2022lora}
that is preconfigured for the FLUX.1-dev model~\cite{aitoolkitLoRA} as our starting point
and provided it with a trigger word, image inputs, and associated preconditions as annotations.
We did not optimize the fine-tuning process via hyperparameter tuning, so the results
reported below represent a lower bound on what might be achieved. 

We trained two sets of LoRAs.
For each requirement, $R$, we filtered the training data to extract the
images that met the requirement precondition to train a LoRA, $L_R$;
we refer to these as \textit{per precondition} LoRA. 
We add a second LoRA for CelebA-HQ trained based on human labels, $L^H_R$.
Finally, as a baseline, we used all of the training data with glossary term annotations
to train a LoRA, $L_{All}$.  
All LoRAs were trained using a \textit{trigger phrase}
that was included in the text associated with images.
All training used a single NVIDIA-A100 GPU with 80 GB memory and the
average LoRA training times were: 52.1 minutes for MNIST, 76.7 minutes for CelebA-HQ, and 82.5 minutes for SGSM.

\subsubsection{Baselines}\label{sssec: baselines}
\technique is the first method to generate test inputs that target
semantic feature-based requirements, so there are no truly comparable baselines.
We can, however, assess how well state-of-the-art test generation techniques that \textbf{do not} target requirements
perform in the \technique context.  To do this we selected two recently published
test generation method as baselines.  The first baseline generates inputs using rotation, blur, brightness and translation image transformations~\cite{demir2024test}, and the second is DeepHyperion-CS~\cite{DeepHyperion-CS} that uses its own set of feature-level variations to generate test inputs. 

\subsubsection{Metrics}
\label{subsec:metrics}
RQ1 requires a metric to quantify the consistency
of generated images with preconditions.  
To measure this, we trained a set of binary classifiers for glossary terms for each dataset, which
we term a \textit{glossary term classifier} (GTC).
Each GTC begins as a pre-trained ResNet-101 model, specifically
\texttt{torchvision.models.resnet101}, that is modified by adding a binary classification head. 
We develop separate training sets for each GTC by selecting positive and negative training samples and balancing them;
full details of this process are available~\cite{rbt4dnn}.
\torepo{
To train a Glossary Term Classifier (GTC), we first held out test data from the training data. 
To do that, we first computed the set, $D = [D_1, D_2,..., D_l]$, 
where $D_i$ is a set satisfying the requirement $i$. 
We also computed $\overline{D} = [\overline{D_1}, \overline{D_2},..., \overline{D_l}]$, where $\overline{D_i}$ is a set that does not satisfy requirement $i$. 
To construct the test set, we sorted $D$ and inserted $r$\% from the smallest set of $D$ into the test set.
Then, we moved to the next smaller set and inserted the data absent in the test set and previously considered sets. While inserting data from $D_i$, we also ensured that the amount of the data in the test set satisfying requirement $i$ is not more than $r$\% of $D_i$. 
We repeated the same procedure for $\overline{D}$ with an additional checking that $D$ did not have the inserted data. 

For each glossary term, we split the training data to include an equal number of randomly chosen inputs with and without the glossary term. 
We randomly held out 10\% of the data with and without the glossary term for the validation set. 
Then, we trained the GTC model over the filtered train set and validated it using the validation set.
}
Across all GTCs for all datasets the mean test accuracy was 94.5\%; a notable outlier was requirement C4 with test accuracy of 70.8\% which is discussed below. 

We used GTCs in both RQ1 and RQ3.  In RQ3, they were used to
compute the relative entropy over glossary terms to characterize the \textit{feature diversity}
of generated tests compared to training data.
More specifically, we reported the Jensen-Shannon (JS) divergence~\cite{menendez1997jensen} in the distribution of glossary terms
between generated images for a requirement precondition and training data filtered
by that precondition. The JS divergence is a more refined version of the Kullback-Leibler (KL) divergence that can handle disjoint sets of glossary terms.

To evaluate RQ2, we need a metric to judge how realistic an image is relative to
the training data. We explored the use of the FID (Fréchet Inception Distance)~\cite{heusel2017gans}, but found that it is very inaccurate
for small sample sizes, which we have for requirements that
describe rare glossary term combinations. 
Consequently, we used the related, but sample efficient and robust  
KID (Kernel Inception Distance)~\cite{binkowski2018demystifying} 
which provides reliable scores even with small sample sizes. 

For RQ4-RQ6, we measured fault-detection effectiveness and analyzed an LC's behavior by running the
selected LCs on  generated test inputs for a requirement and reported the percentage
of outputs that satisfied the postcondition.  
For RQ5, we also estimated the false positive rate for
test failures by performing multi-assessor human evaluation of random samples.

\subsection{Results and Analysis: Quality of Generated Tests}

\subsubsection{\rq{1}: How consistent are \technique generated images with  requirements?}
\label{sssec:rq1}
\begin{figure}[t!]
    \centering
    \includegraphics[width=\textwidth]{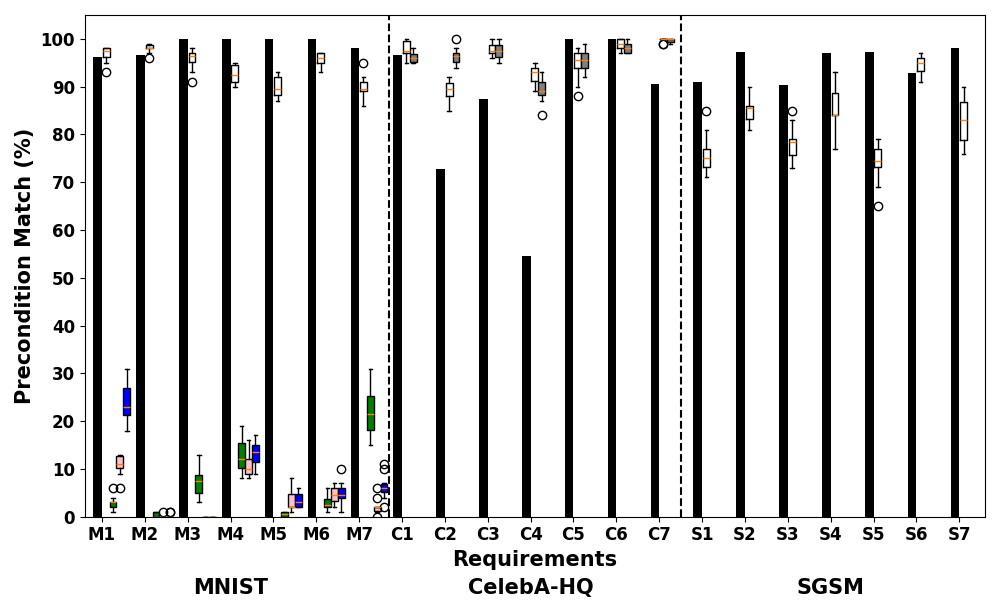}%
    \caption{Percentage of images matching preconditions as judged by GTCs. All required features (AND) must match labels. The box plots show data for:
    $L_R$ (white), $L_{All}$ (green),
    Image transform baseline (blue), DeepHyperion (pink), and $L_R^H$ (gray).}
    \label{fig:rq1}
\end{figure}

To address this question, we passed generated images through the GTCs mentioned in a requirement precondition and then computed
the logical combination of GTC outcomes in the precondition to determine a \textit{precondition match} for the image.
GTCs are imperfect classifiers and this can introduce noise into our measurement. To account for this we computed, for each precondition, 
\ignore{
a subset of the held-out test data that is balanced with respect to positive and negative precondition outcomes.}
a subset of the held-out test data with positive precondition outcomes and
reported the percentage of samples that match as black bars in Figure~\ref{fig:rq1}.  A black bar for a precondition that falls below 100\% indicates that at least one of the GTCs for that precondition is inaccurate.
The black bars average 98.69\% for MNIST,
85.99\% for CelebA-HQ, and and 94.84\% for SGSM.

\begin{figure}[t!]
    \centering
    \includegraphics[width=0.9\textwidth]{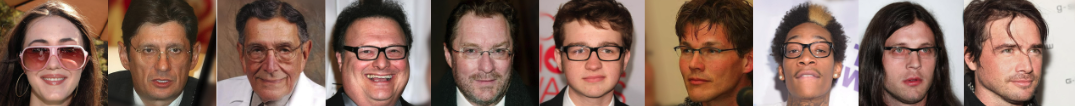}%
    \caption{Held-out CelebA-HQ test samples whose glossary term labels indicate a C4 precondition match, but where GTC inferences resulted in a mismatch.}
    \label{fig:rq1-gtc-c4}
\end{figure}

We note that some GTCs for CelebA-HQ are inaccurate -- particularly
for requirements C4 and C2.
We investigated the low match value for C4, ``The person is wearing eyeglasses and has wavy hair'', by analyzing the 10 images from
the test dataset that were labeled in such a way that C4 should be true, but
where GTC inferences found C4 to be false -- see Figure~\ref{fig:rq1-gtc-c4}.
A  manual analysis of these images found: 1 with no glasses and 4 that do not have wavy hair.
These are errors in the human labeled glossary terms in the dataset. 
Such labeling errors explain the low test accuracy of 70.8\% for the ``wavy hair'' GTC, and since this GTC is used in the C4 precondition, it explains why
the precondition match is so low for the training data, 54.6\%.
A similar analysis explains C2's performance on test data as one of its GTC's
had the second lowest test accuracy of 83.4\%, also due to mislabeling.
Labeling errors can happen, but we conjecture that combinations of glossary
terms in a precondition mitigates the impact as we discuss below.

For each LoRA and each requirement, we generated 100 images
and computed the precondition match percentage.  We repeated
this 10 times and report statistics through box plots
in Figure~\ref{fig:rq1}.

For MNIST, figure~\ref{fig:rq1} shows, from left to right, results for two LoRA models: $L_R$ (white) and $L_{All}$ (green), and the two baseline techniques: DeepHyperion (pink) and Image transformation (blue).
We can draw two conclusions from these data.
First, combining multiple requirements into a single LoRA, 
$L_{All}$, negatively impacts the ability to match preconditions.  
Second, the baselines techniques -- pink and blue -- are unable to match
preconditions at a high rate.
These techniques are not
designed to target a precondition, so their poor performance is not surprising,
but these data substantiate that existing approaches are not effective in the
\technique context. 
Given their poor performance, we do not study $L_{All}$ or the
two baselines in the rest of this evaluation.

\begin{figure}[t!]
    \centering
    \begin{minipage}{0.9\textwidth}
        \includegraphics[width=0.1\linewidth]{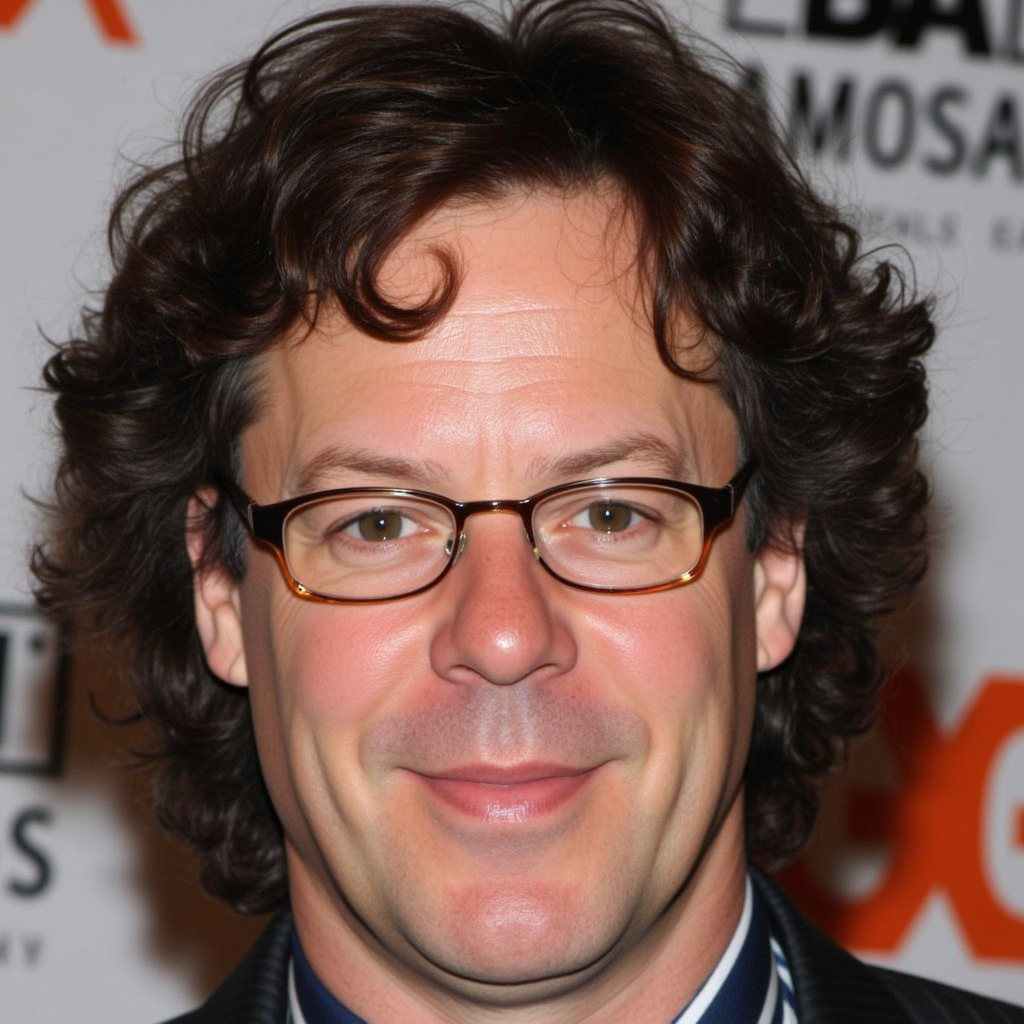}  
        \includegraphics[width=0.1\linewidth]{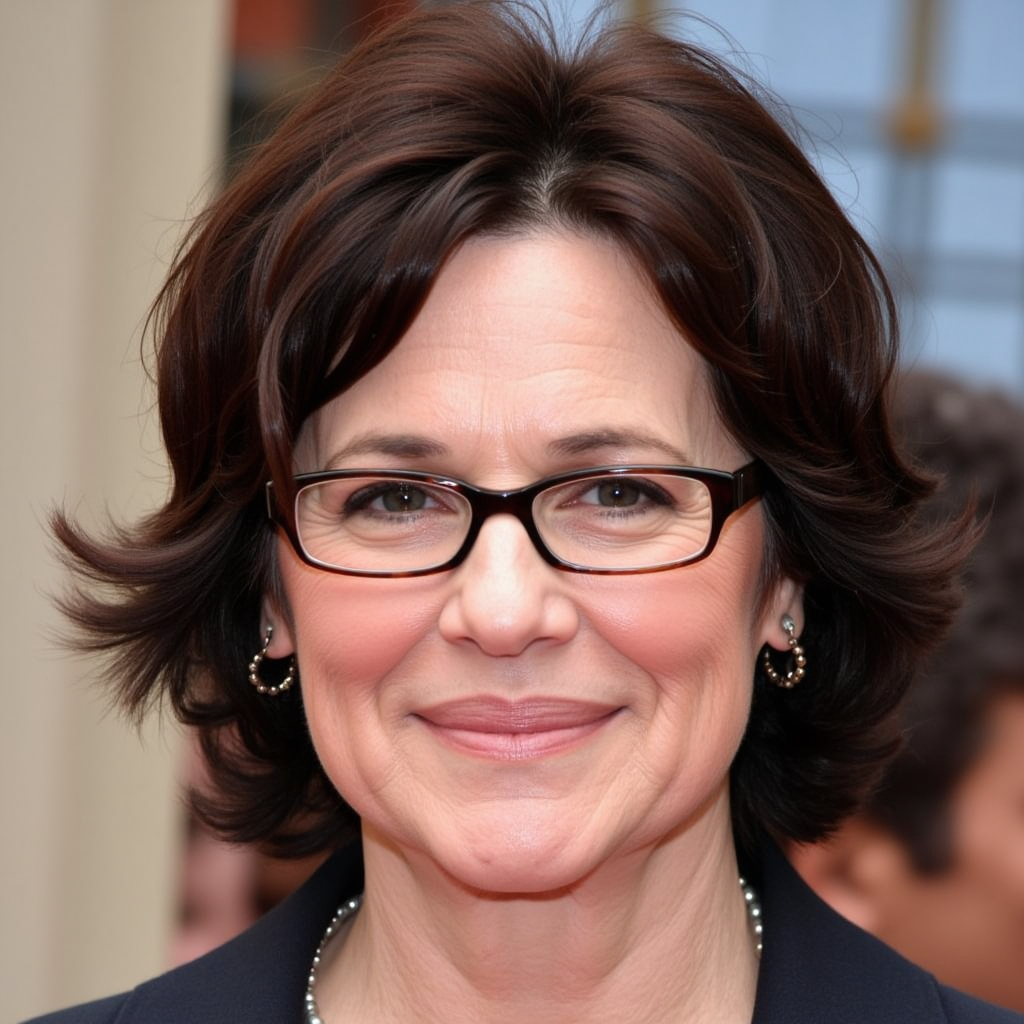}
        \includegraphics[width=0.1\linewidth]{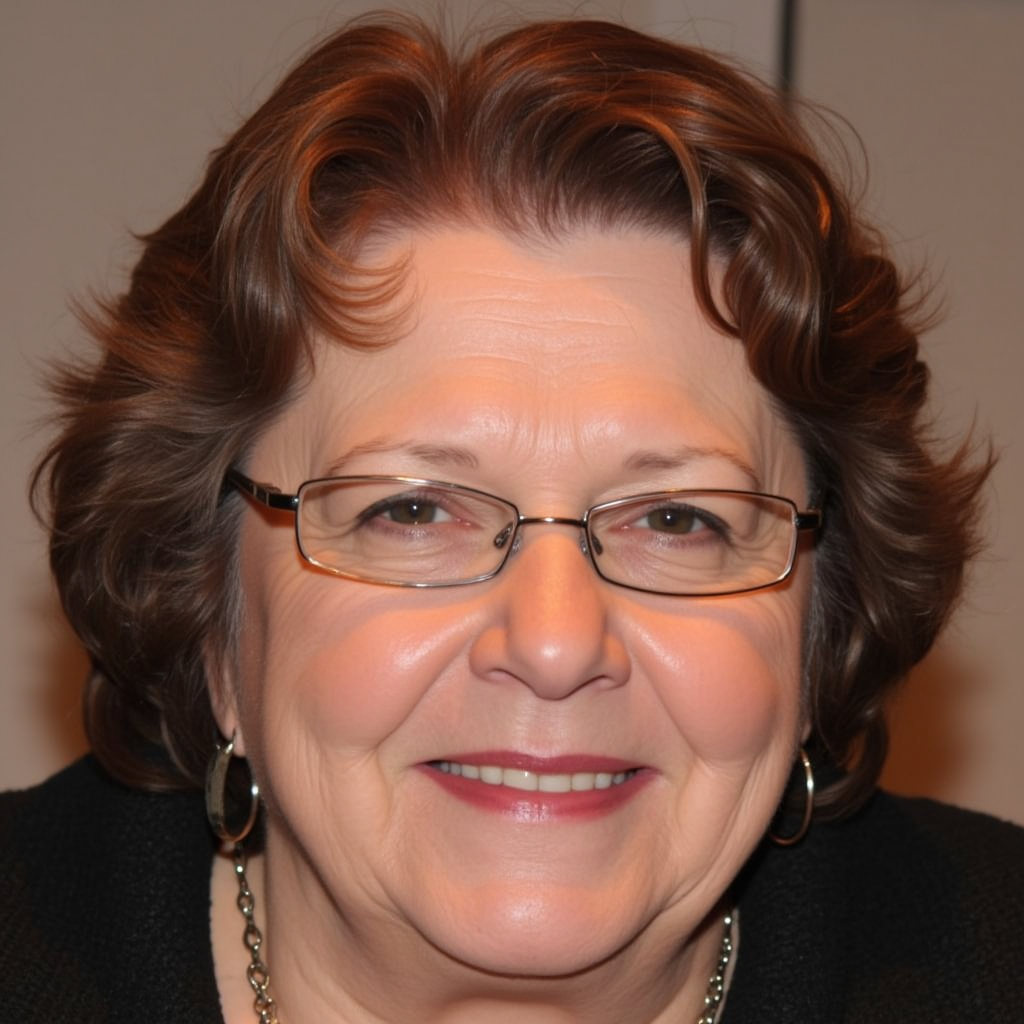}   
        \includegraphics[width=0.1\linewidth]{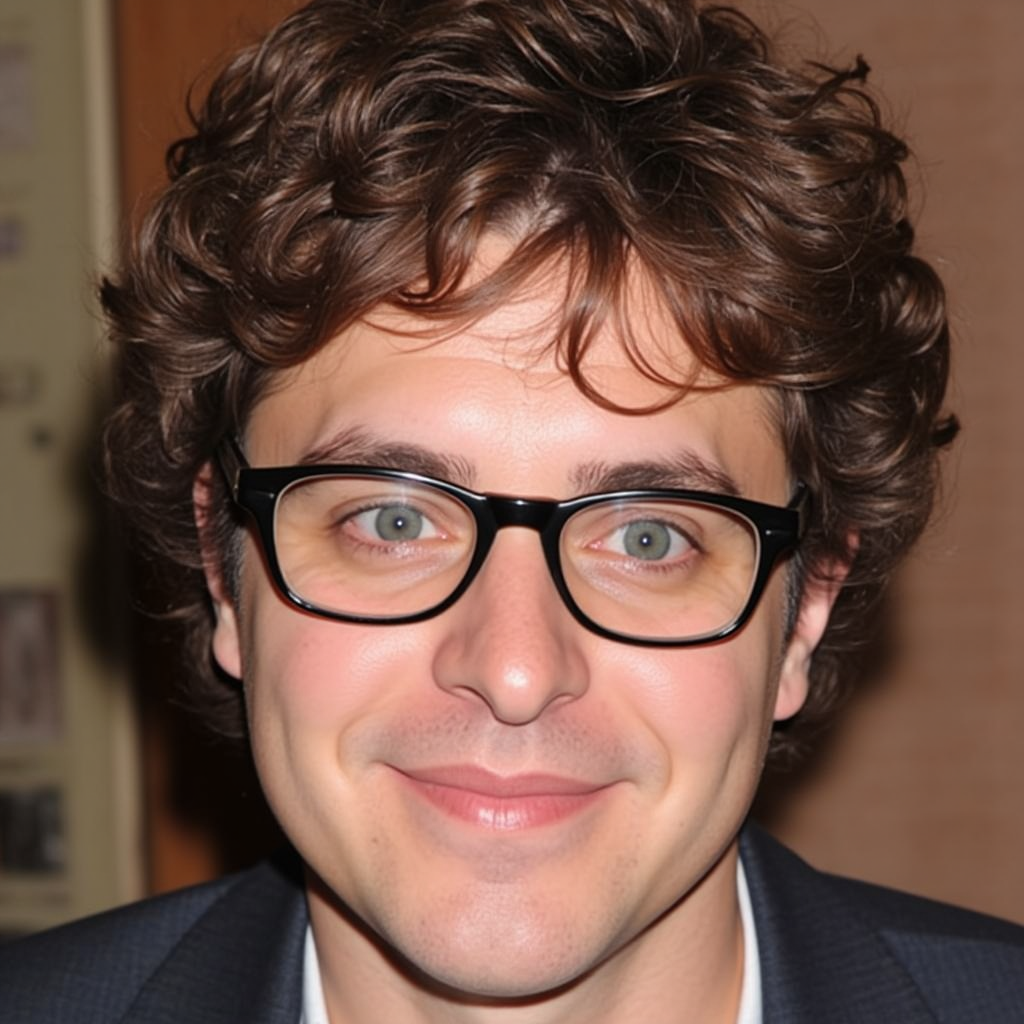} 
        \hfill
        \includegraphics[width=0.1\linewidth]{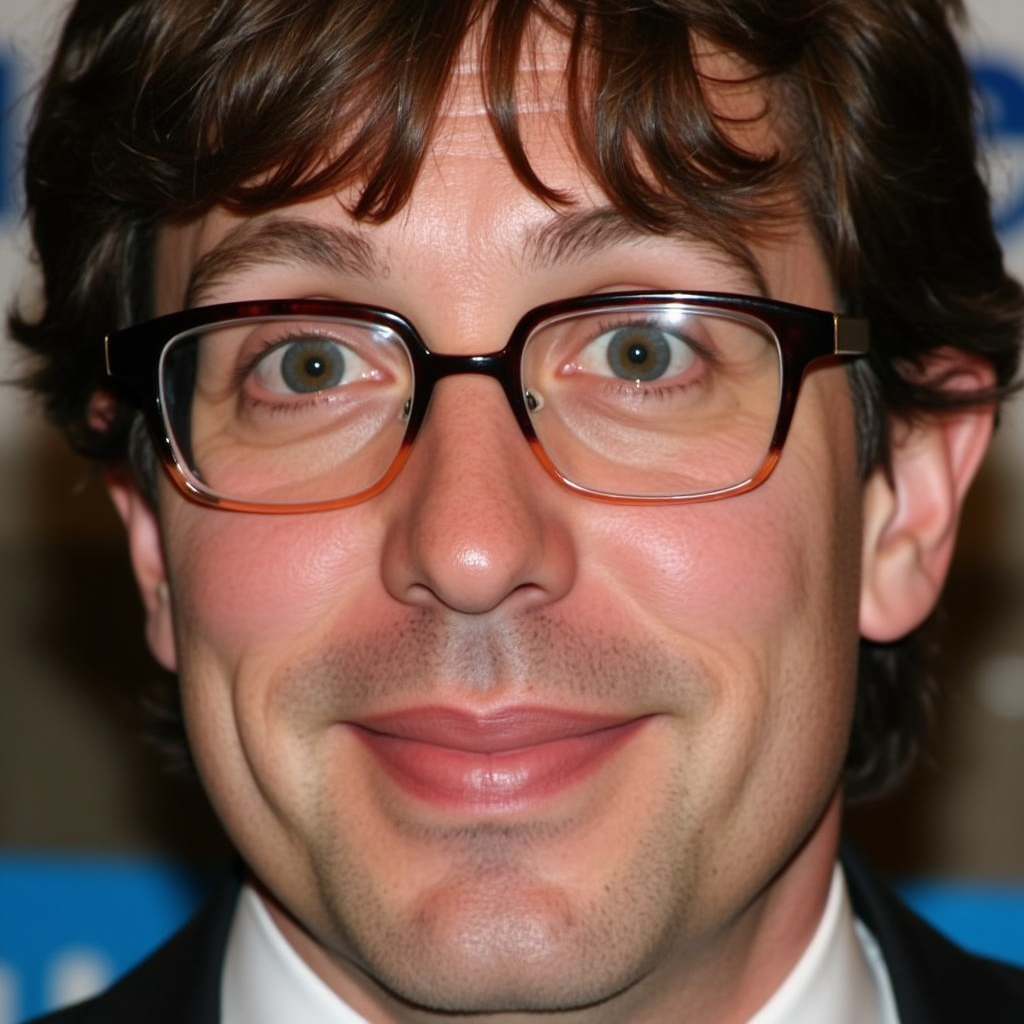}  
        \includegraphics[width=0.1\linewidth]{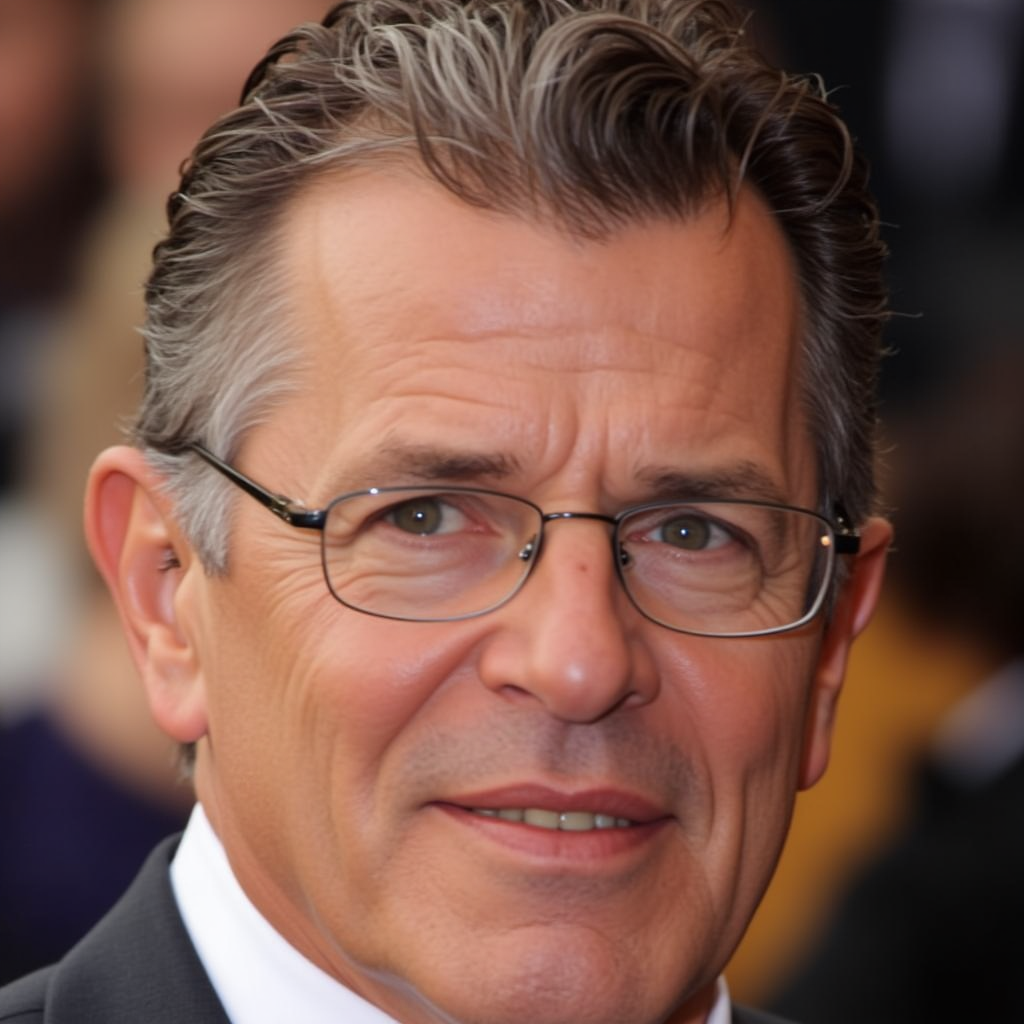} 
        \includegraphics[width=0.1\linewidth]{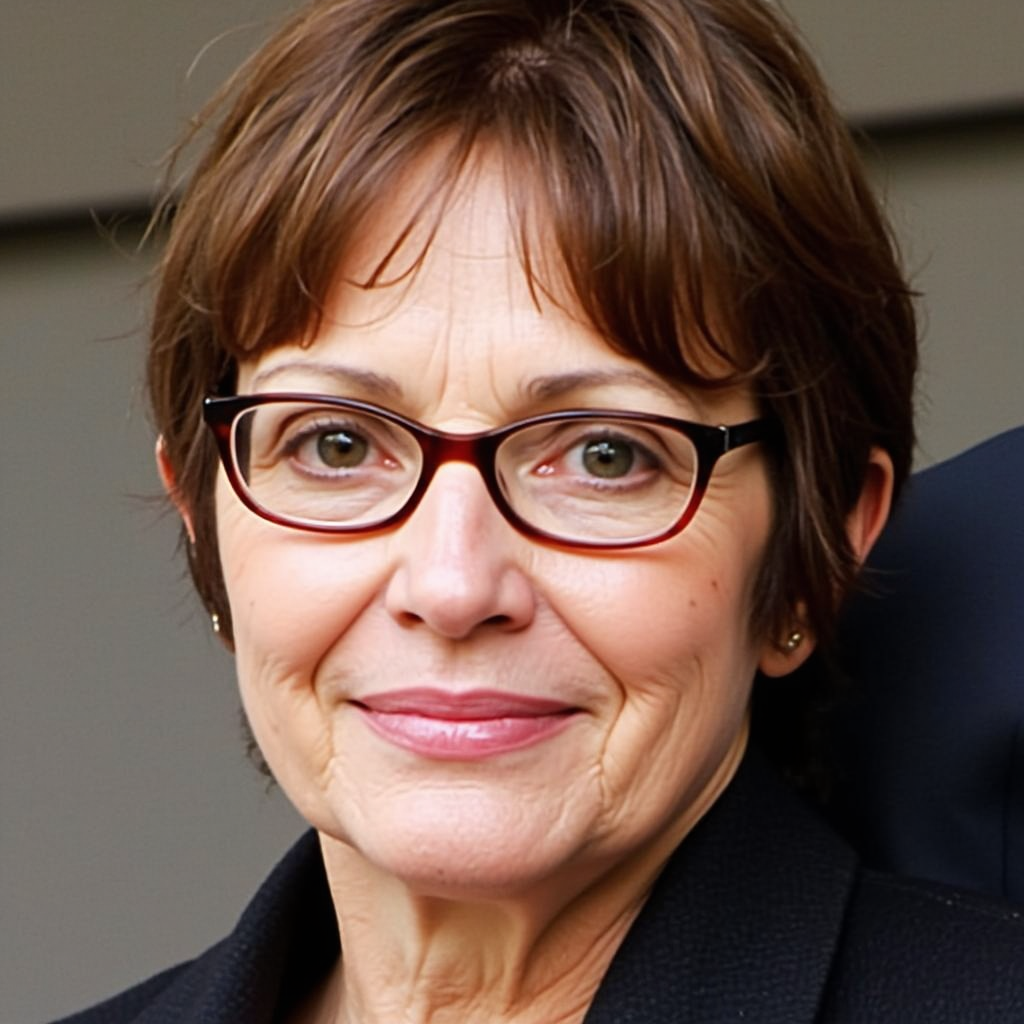} 
        \includegraphics[width=0.1\linewidth]{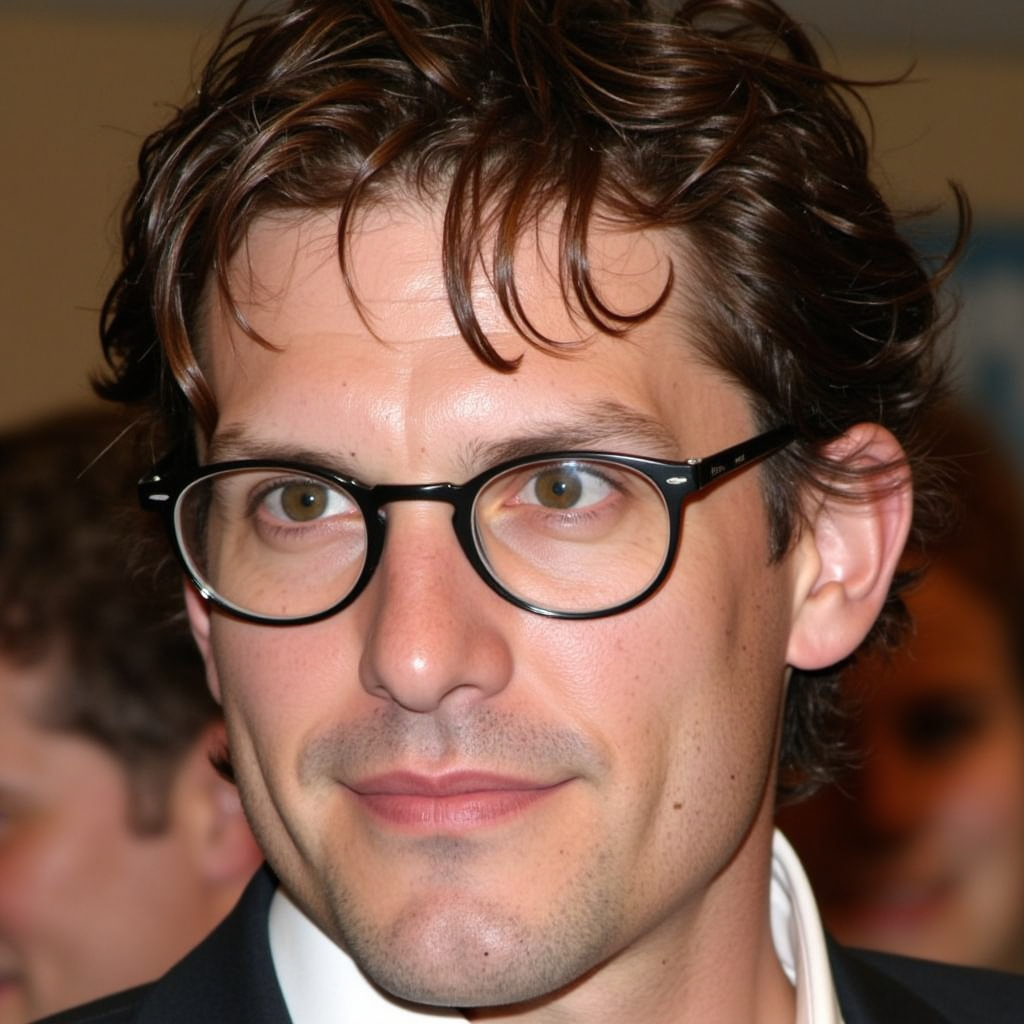}
    \end{minipage}

    \caption{Random samples of generated images for C4 precondition.  The leftmost 4 images are detected as matching the precondition and the rightmost 4 are detected as not matching.}
    \label{fig:rq1_celebahq_samples}
\end{figure}

For CelebA-HQ, we used two LoRA models, $L_R$ (white) trained using VQA generated labels and $L^H_R$ (gray) trained using human labels.
$L^H_R$ generated images have precondition match values between 84\% and 100\% with an average of 96.1\% across the requirements, while $L_R$ spanned similar range with an average of 95.8\%.
Requirement C4 and C2 had the lowest precondition match for $L^H_R$ and $L_R$, respectively, with 84\% and 85\% precondition match. 
Recall that for these requirements, we identified labeling errors as a source of low GTC accuracy.
We conjecture that even in the presence of labeling errors, the trained GTC still learn to extract salient features -- the curly hair in the absence of glasses and the glasses in the absence of curly hair. 
In this way, the GTC learn some, if not all, of the features in the precondition from each training sample.
Similarly, we conjecture that the LoRA is also able to extract the salient features, and also leverage the original Flux model prior knowledge about these features given that it has been trained with a wide variety of data and text.
To substantiate this, we generated randomly selected images generated for C4,
Figure~\ref{fig:rq1_celebahq_samples},
that matched the precondition (on the left)
and mismatched (on the right) for C4.
The matched images clearly exhibit the precondition characteristics, but at least half of the mismatched images do as well.
This suggests a degree of robustness in the consistency of generated images
relative to glossary terms errors.

\begin{figure}[t!]
    \begin{minipage}{\textwidth}
        \centering

        \includegraphics[width=0.245\linewidth]{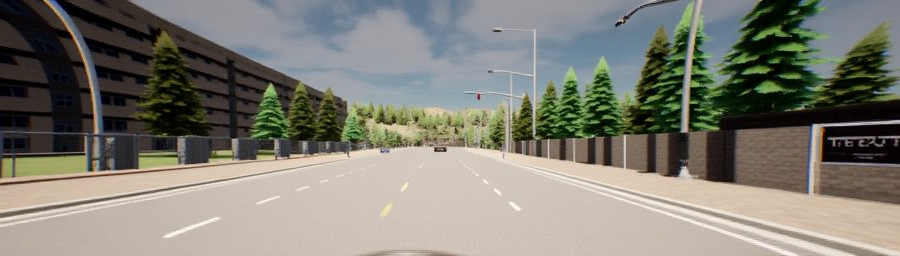}  
        \includegraphics[width=0.245\linewidth]{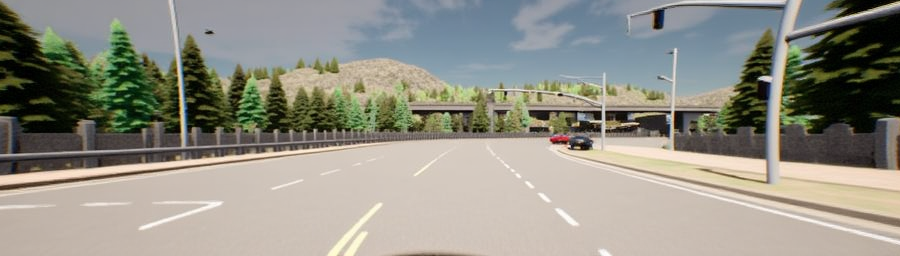}
        \includegraphics[width=0.245\linewidth]{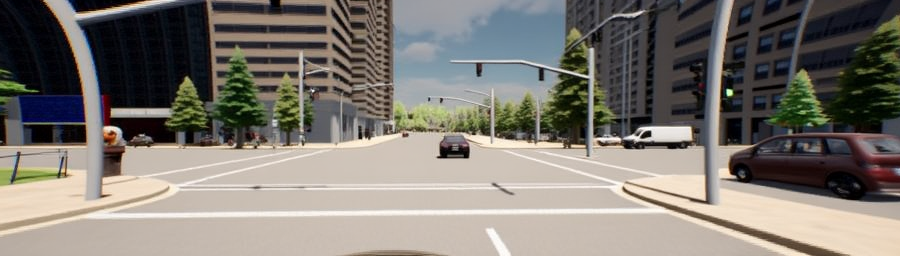}   
        \includegraphics[width=0.245\linewidth]{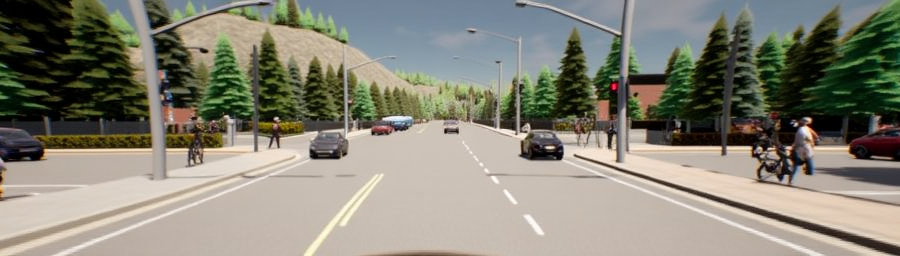}
       
        \includegraphics[width=0.245\linewidth]{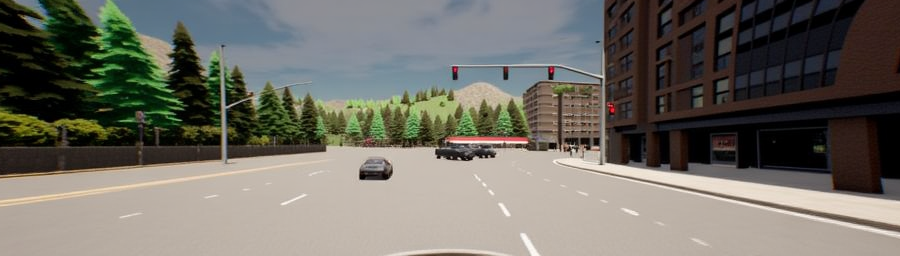}  
        \includegraphics[width=0.245\linewidth]{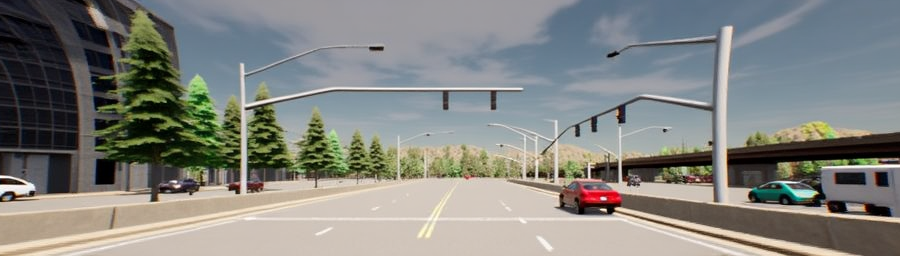}
        \includegraphics[width=0.245\linewidth]{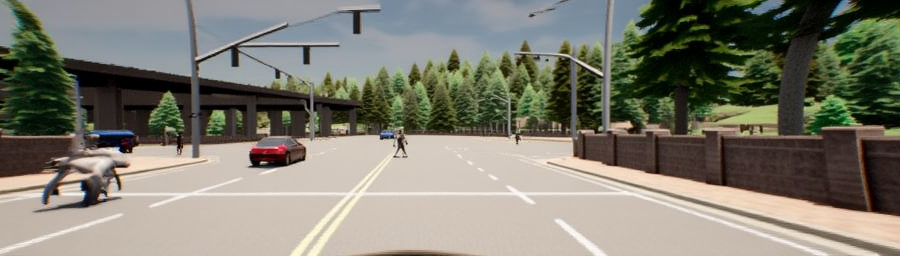}   
        \includegraphics[width=0.245\linewidth]{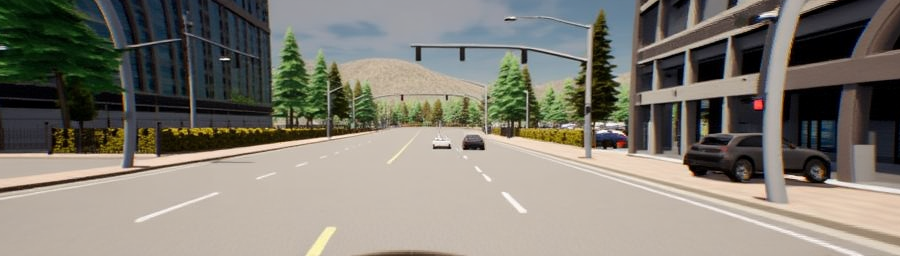}
    \end{minipage}
    \caption{Random samples of generated images for S5 precondition.   The top 4 images are detected as matching the precondition and the bottom 4 are detected as not matching.}
    \label{fig:rq1_sgsm_samples}
\end{figure}

For SGSM, $L_R$ generated images have precondition match values between 65\% and 97\% with an average of 82.3\% across the requirements.
We investigated requirement S5 which had the lowest precondition match, by
exploring sample images, in Figure~\ref{fig:rq1_sgsm_samples}, that were classified as matching (top) and mismatching (bottom) the precondition
``The ego is in the leftmost lane and is not in a intersection''.
None of the images shows the ego vehicle in the intersection, and all but one (the lower left image in Figure~\ref{fig:rq1_sgsm_samples}) clearly show the ego vehicle in
the left lane. The lack of lane markings in that one image makes it challenging to determine which is the left lane.
This analysis suggests that $L_R$ is capable of generating images that satisfy the 
precondition -- perhaps even at a higher rate than is reported by
the precondition match percentage.


\noindent\fbox{
    \parbox{0.97\textwidth}{
       \textbf{\rq{1} Finding: Across 3 datasets and 21 requirements, per-precondition LoRA generated inputs were consistent with preconditions 92.1\% of time, which significantly outperforms two state-of-the-art baselines.
       }
    }
}

\subsubsection{\rq{2}: How realistic are the \technique generated images?}
\label{sssec:rq2}
To address this question, we computed the KID score using~\cite{obukhov2020torchfidelity} to compare the generated images for each requirement to the images in the training dataset that meet the requirement.  To supplement this quantitative assessment we show
image panels in Figures~\ref{fig:rq1_celebahq_samples}, \ref{fig:rq1_sgsm_samples}, and Table~\ref{tab:violationsamples} for qualitative
assessment.

We compared the KID scores for images generated using $L_R$, for each
requirement, and for images generated from Flux using a prompt
that includes the precondition (both $L_R$ and Flux used identical prompts).  
Both of these sets of images had their KID score
computed relative to 
the subset of the training data satisfying the precondition.
For CelebA-HQ, we also computed KID for Flux and $L_R^H$ which used only human labels to filter the data.

As shown in Table~\ref{tab:rq2}, $L_R$ generated images for MNIST (M1-M6) and SGSM (S1-S7) show significant reduction in the KID score relative to Flux model -- an average of 77.09\% improvement.
For CelebA-HQ, the data show comparable performance for $L_R$ and $L_R^H$,
but both show a smaller reduction relative to Flux,
with an average of 12.9\% and 27\%, respectively.  We conjecture that since Flux was trained on an enormous amount of data, including large numbers of real human faces, it is already quite good at generating realistic
human faces given a prompt like ``A close head shot of a person ...'', where
```...''' is replaced with the SNL precondition. 
This is consistent with the fact that the  KID(F) scores for CelebA-HQ (C1-C7) are substantially lower than for MNIST and SGSM.  

Published research has reported KID scores for the MNIST and CelebA datasets, e.g., \cite{binkowski2018demystifying}, that are lower than the scores
we report here for $L_R$ generated images.
We note, however, KID is sensitive to sample size as depicted in Figure 1 in \cite{binkowski2018demystifying} where sample sizes below 500 can increase the KID score. 
Because we are filtering the dataset based on preconditions, for a number of
our requirements, including those for MNIST and CelebA-HQ, we have fewer
than 300 samples.  We conjecture that this is a contributing factor to 
higher KID scores for $L_R$ than one might expect from looking at random samples, e.g., Figures~\ref{fig:rq1_celebahq_samples} and~\ref{fig:rq1_sgsm_samples}.  

\begin{table}[t]
    \small
    \centering
    \resizebox{0.8\textwidth}{!}{
    \begin{tabular}{|c|p{1.2cm}|p{1.2cm}||c|p{1.2cm}|p{1.2cm}|p{1.2cm}||c|p{1.2cm}|p{1.2cm}|}
    \hline
    ID & KID(F) Mean & KID($L_R$) Mean & ID & KID(F) Mean & KID($L_R^H$) Mean & KID($L_R$) Mean &  ID & KID(F) Mean & KID($L_R$) Mean\\
    \hline
    M1 & 0.193 & 0.069 & C1 & 0.123 & 0.072 & 0.097 &  S1 & 0.149 & 0.036\\
    \hline
    M2 & 0.355  & 0.049  & C2 & 0.097  & 0.086  & 0.108  &  S2 & 0.242  & 0.038 \\
    \hline
    M3 & 0.308  & 0.039  & C3 & 0.117  & 0.088  & 0.085  &  S3 & 0.312  & 0.035 \\
    \hline
    M4 & 0.167  & 0.061  & C4 & 0.078  & 0.057  & 0.084  &  S4 & 0.123  & 0.041 \\
    \hline
    M5 & 0.181  & 0.023  & C5 & 0.058  & 0.049  & 0.058  &  S5 & 0.113  & 0.048 \\
    \hline
    M6 & 0.226  & 0.019  & C6 & 0.094  & 0.058  & 0.057  &  S6 & 0.116  & 0.030 \\
    \hline
    M7 & 0.138  & 0.036  & C7 & 0.110  &  0.076 &  0.086 &  S7 & 0.177  & 0.039 \\
    \hline
    \end{tabular}}
    \caption{Mean KID score for generated images over each precondition for each dataset for Flux (F) model with base prompt and $L_R$; $L^H_R$ shown for CelebA-HQ for comparison. The maximum standard deviation across these data is 0.010.}
\label{tab:rq2}
\end{table}

\noindent\fbox{%
    \parbox{0.97\linewidth}{%
       \textbf{\rq{2} Finding: For datasets that are not well-represented in pre-trained generative models, like MNIST and SGSM, $L_R$ generated images are substantially more realistic than those produced just by prompting.}}
}

\subsubsection{\rq{3}:  How diverse are the \technique generated images?}
\label{sssec:rq3}
When a precondition holds in an image, this may impact the presence, or absence,
of glossary terms not mentioned in the precondition.  For example, if a digit is ``very right leaning'' then it cannot also 
be ``left leaning''.
Correlations like these will impact the diversity of training data when filtered
by precondition and it is the diversity of such filtered data that we use as
a baseline for judging the diversity of generated test inputs.

\begin{table}[t]
    \small
    \centering
    \resizebox{0.8\textwidth}{!}{
    \begin{tabular}{|c|p{1.1cm}|p{1.1cm}||c|p{1.1cm}|p{1.1cm}|p{1.2cm}||c|p{1.1cm}|p{1.1cm}|}
    \hline
    ID & JS(F) & JS($L_R$) & ID & JS(F) & JS($L_R^H$) & JS($L_R$) &  ID & JS(F) & JS($L_R$) \\
    \hline
    M1 & 0.18136 & 0.08371 & C1 & 0.15683 & 0.10456 & 0.12538 &  S1 & 0.07089 & 0.04284 \\
    \hline
    M2 & 0.24824 & 0.00889 & C2 & 0.13823 & 0.15521 & 0.12566 &  S2 & 0.24443 & 0.03639 \\
    \hline
    M3 & 0.22337 & 0.00665 & C3 & 0.03582 & 0.02451 & 0.03712 &  S3 & 0.21668 & 0.04772 \\
    \hline
    M4 & 0.14558 & 0.01263 & C4 & 0.13212 & 0.07858 & 0.12365 & S4 & 0.22287 & 0.02311 \\
    \hline
    M5 & 0.19907 & 0.02824 & C5 & 0.06486 & 0.04324 & 0.04034 &  S5 & 0.13476 & 0.03577 \\
    \hline
    M6 & 0.16753 & 0.00681 & C6 & 0.08485 & 0.01844 & 0.01543 &  S6 & 0.18629 & 0.03796 \\
    \hline
    M7 & 0.07018 & 0.02358 & C7 & 0.19198 & 0.10294 & 0.11395 & S7 & 0.29609  & 0.02443 \\
    \hline
    \end{tabular}}
    \caption{Comparing Jensen-Shannon Divergence (JS) between training images that meet a requirement and generated Images from Flux (F) and $L_R$, and $L_R^H$ for CelebA-HQ requirement.}
\label{tab:rq3}
\end{table}

We measured the relative diversity of preconditions for generated and training data using JS divergence over sets of glossary terms using 1000 generated images from each $L_R$ and Flux for all requirements.

Table~\ref{tab:rq3} shows the JS divergence between the training and generated images. The JS divergence of the $L_R$ model for each requirement is close to zero, indicating that the generated images follow the training  diversity over the glossary terms. 
For MNIST and SGSM requirements the JS score for Flux is 2 to 20 times higher
than for $L_R$.
For CelebA-HQ, Flux is closer to the training distribution than MNIST and SGSM and
both $L_R$ and $L_R^H$ are closer to Flux.
As discussed above in \rq{2}, we 
conjecture that this is because Flux has been trained on a large number
of human faces and has learned a good representation of that diversity.

Across the requirements, the JS scores indicate that $L_R$ images for MNIST and SGSM requirements are more three times
closer, on average, to the training data than Flux generated images.
For the CelebA-HQ, the improvement is more modest at 32\%, but in a more favorable
setting for Flux.

\noindent\fbox{
    \parbox{0.97\textwidth}{
       \textbf{\rq{3} Finding: The feature diversity in $L_R$ generated images is highly consistent with the diversity of training images.}}
}
\subsection{Results and Analysis: 
Applicability of Generated Tests}
\label{ssec:applicability}
The following research questions address the limitation of existing techniques in assessing model confidence in requirement-specific cases and how \technique overcomes it. They also assess the effectiveness of \technique generated tests in revealing requirement-specific faults and unobserved model behavior.
For this study, we generated inputs for the requirements from Table~\ref{tab:requirements} using a precondition specific $L_R$, manually encoded the postcondition
of the requirement to define a test oracle, and then executed the LC on 
each input checking the oracle. 
We generated 1000 tests for each precondition and measured the number of tests that fail the corresponding postcondition. 
This experiment was repeated 10 times and the results are presented as box plots in Figure~\ref{fig:rq4} and Figure~\ref{fig:case_study}.

\subsubsection{\rq{4}: How does \technique compare to baselines in assessing requirement-specific model behavior?}
\label{sssec:rq4}
The purpose of testing an LC is to assess its ability to perform on unseen data and ensure that the model aligns with specific application requirements. A high test pass rate demonstrates the LC's ability to perform well in cases that the tests represent. 
It is essential, however, to understand what those tests represent:
if they do not meet the requirement then a test pass -- or a test
failure -- provide no information about the LC's behavior relative
to the requirement.

To investigate the performance of existing techniques in identifying requirement-specific faults, we selected two recent test methods, Deephyperion and Image Transform-generated tests, and applied them 
on the MNIST dataset.
The pass results are presented in the top-row of Figure~\ref{fig:rq4}.
The data show very high-pass rates for all methods with a few outliers:
\technique (left panel) on M3, and ImageTransform on M1 and M6.
However, as we showed in Figure~\ref{fig:rq1} both DeepHyperion and
and Image Transformation have very low precondition match percentages
which means that these high pass rates are not actually evaluating the LC on the requirement precondition.

To investigate further, we selected random samples from passing and failing tests for each test generation technique for requirement M2 -- \textbf{the digit is a 3 and is very thick}.
Table~\ref{tab:diff_techniques} shows the samples and it is clear
that neither DeepHyperion nor Image Transformation generates thick digits.
This means that is likely that neither the pass nor the fail results are 
meaningful for the precondition.  Consequently, the high passing rates should not be interpreted as evidence that LC’s behavior is aligned with requirement M2.

\noindent\fbox{
    \parbox{0.97\textwidth}{
       \textbf{\rq{4} Finding: The tests generated by \technique provide a better basis for establishing confidence in model behavior, as they are explicitly guided by requirements, unlike existing techniques that do not adapt test generation to requirement preconditions.}
}}
\begin{table}[t!]
    \centering
    \begin{tabular}{|c|c|c|} 
    \hline
    Technique & Passed Tests & Failed Tests \\
        \hline
        Deephyperion & 
        \begin{minipage}{0.35\textwidth}
            \centering
            \vspace{1pt}
        \includegraphics[width = 0.9\linewidth]{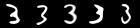} 
        \vspace{1pt}
        \end{minipage} &
        \begin{minipage}{0.35\textwidth}
            \centering
            \vspace{1pt}
        \includegraphics[width=0.9\linewidth]{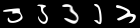} 
        \vspace{1pt}
        \end{minipage} \\ 
        \hline
        
        Image Transform & 
        \begin{minipage}{0.35\textwidth}
            \centering
            \vspace{1pt}
            \includegraphics[width=0.9\linewidth]{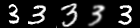}
            \vspace{1pt}
        \end{minipage} &
        \begin{minipage}{0.35\textwidth}
            \centering 
            \vspace{1pt}
        \includegraphics[width=0.9\linewidth]{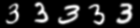} 
        \vspace{1pt}
        \end{minipage} \\
        \hline

        RBT4DNN & 
        \begin{minipage}{0.35\textwidth}
            \centering
            \vspace{1pt}
            \includegraphics[width=0.9\linewidth]{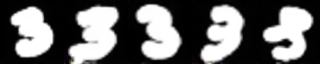}
            \vspace{1pt}
        \end{minipage} &
        \begin{minipage}{0.35\textwidth}
            \centering 
            \vspace{1pt}
        \includegraphics[width=0.9\linewidth]{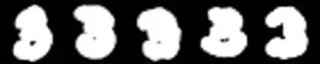} 
        \vspace{1pt}
        \end{minipage} \\
        \hline
    \end{tabular}
    \caption{Random passing and failing test samples from Deephyperion, Image Transform and \technique  for M2 : \textbf{the digit is a 3 and is very thick.}}
    \label{tab:diff_techniques}
\end{table}

\subsubsection{\rq{5}:  How effective are \technique generated tests in revealing faults?}
\label{sssec:rq5}
\begin{figure}[t!]
    \centering
    \includegraphics[width=\linewidth]{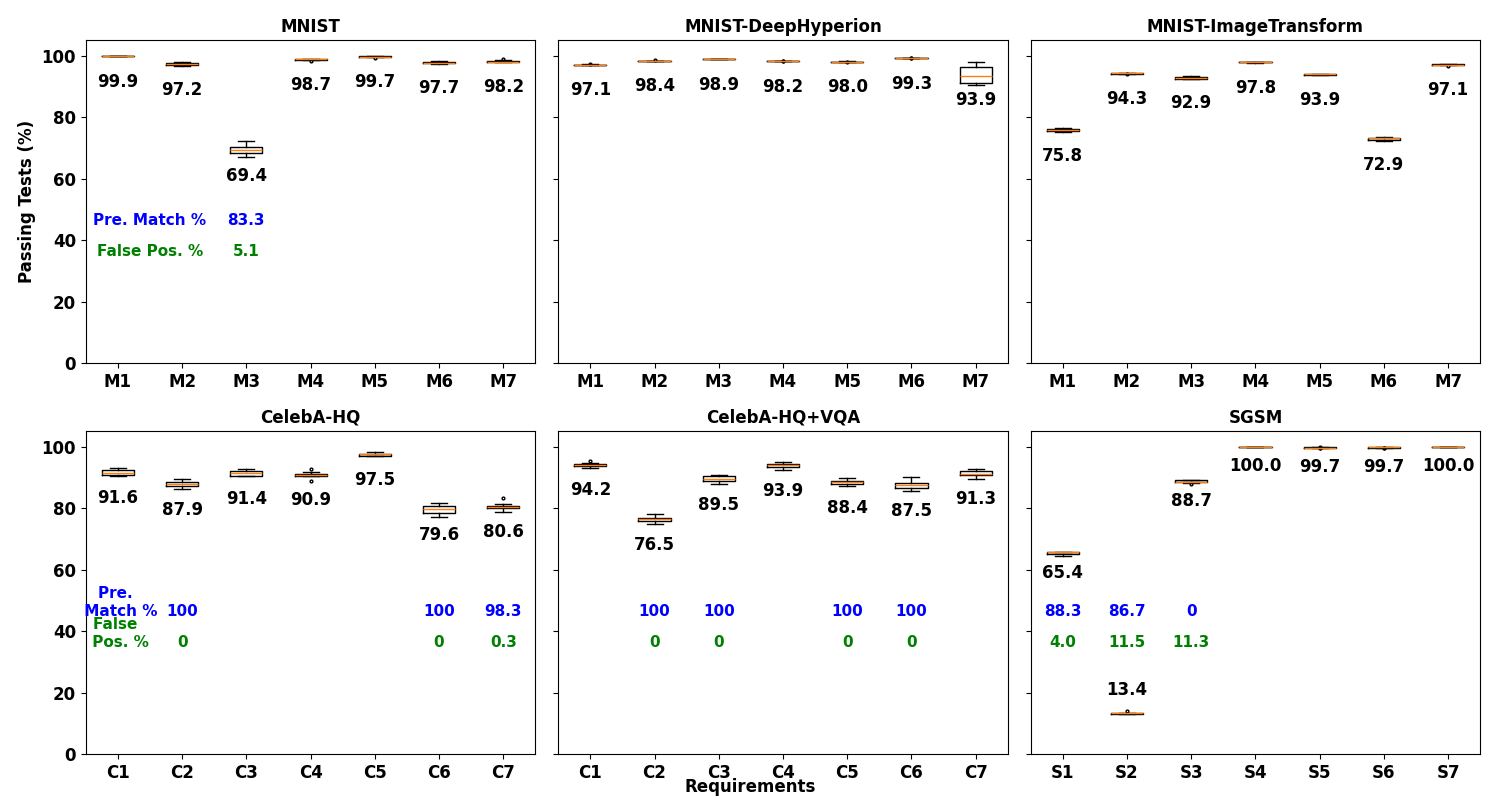}
    \vspace{-6mm} 
    \caption{Percentage of generated tests for each requirement that passed the post-condition. The numerical values in black in the plots are the mean values. The numerical values in blue are the percentage of inputs for failing tests that match the precondition and the values in green are the  percentage of false positive test cases for requirements with a pass rate of less than 90\%.}
    \label{fig:rq4}
\end{figure}

This research question explores the effectiveness of \technique in identifying when LCs fail to conform to requirements. 
As shown in Figure~\ref{fig:rq4}, our experiments demonstrate that \technique is capable of detecting postcondition failures across all of the requirements except for S1 and S7.  We discuss the failure cases in more detail below.

The MNIST classifier is more prone to failures for M3, while for the other requirements, more than 97.2\% of the generated test inputs passed. 
For CelebA-HQ, we show two plots: $L^H_R$ which uses human feature labels (left of lower row) and $L_R$ which uses VQA generated feature labels (middle of lower row).  
With human labels, three of the requirements had pass rates below 90\% (C2, C6, and C7), where as there were four such requirements with VQA labels (C2,C3,C5, and C6).  
For SGSM, of the requirements with failures, two had 99.7\% pass rates, but the remaining requirements (S1, S2 and S3), had pass rates below 90\% -- with S1 and S2 significantly below that level.

We explored whether failures can reveal faults in LCs by first looking at the requirements with some of the highest pass rates (C1, C4, S5, and S6). Table~\ref{tab:violationsamples} shows randomly sampled test failures of these requirements. 
Recall that the postcondition for C1 and C4 is that the model should predict that the person in the image is wearing eyeglasses. 
The failed test inputs all represent faults, since the images are
all realistic depictions of people wearing glasses in the style of
CelebA-HQ data.
S5 and S6 require that the model does not output a steering angle to the left, but with different preconditions.  Both the passing and failing tests in the table satisfy those preconditions, so it is reasonable to conclude that the failed tests represent faults.

While it is clear that faults can be detected, we dug deeper
into S5 to explore the set of all 25 failing test cases out
of the 10000 shown.  S5's precondition states that \textbf{The ego is in the leftmost lane and not in a intersection}.
In all failure cases, the ego vehicle was not in an intersection,
but the image depicted that the ego vehicle was approaching
an intersection. 
This commonality among failures suggests that LC developers
might undertake a careful consideration of the scenario where the ego approaches the intersection, even though the LC passes the scenario for some samples (as shown in Table~\ref{tab:violationsamples}). This type of analysis is crucial for understanding LC behavior and the scope for LC improvement. 

We saw in RQ1 that the precondition match percentage was quite high across datasets and requirements in our study, though the SGSM match rates were significantly lower than for MNIST and CelebA-HQ.   While RQ1 isolated the question of precondition 
match,  
large numbers of failing tests fail raise the concern that those failures
are \textit{false positives}, i.e., the generated inputs are unrealistic or do not
match the precondition.
To explore this, we conducted a manual study to estimate the false
positive percentage of \technique-generated tests for the 11 requirements where the pass rate was below 90\%.
For each case, multiple authors independently assessed 30 
randomly sampled test inputs that led to a requirement postcondition failure to determine
both realisticness and precondition  consistency.
Assessors agreed that all images were realistic and they disagreed on
precondition consistency for 1.7\% of the 330 inputs that were analyzed.
We aggregated precondition match data from all assessors to estimate the \textit{precondition match percentage}, $pmp$, which is
 shown in blue in Figure~\ref{fig:rq4}.  Generally
these percentages are high, except for S3 which we discuss below.

A test failure is a false positive if the test fails and the input does not
meet the precondition.   The y-axis in Figure~\ref{fig:rq4} is the \textit{passing test percentage}, $ptp$,
and we can estimate the false positive rate as: $(1 - pmp)*(1 - ptp)$, which is shown in green in the Figure.
Estimated false positive rates are all below 11.5\%, and in many cases much lower.  This suggests that a high-percentage of test failures depict scenarios where LC behavior mismatches requirements.

The case of S2 which had a very high rate of true test failures indicates that the LC's behavior is very inconsistent with the requirement which states that \textbf{If the ego lane is controlled by a red or yellow light, then the LC shall decelerate}.  
Our manual analysis of failing inputs for this case showed a red light controlling the ego
lane, but where that traffic light was a long distance down the road.  
With this feedback a developer might choose to refine the requirement, e.g., adding
a ``within 25 meters'' modifier on the traffic light, to better reflect intended behavior.
This case reveals how \technique offers the potential to 
provide feedback on requirements.

While S1 and S3 have very different pass rates and different $pmp$ values for the sampled failing tests, they share a common semantic feature in their preconditions: \textbf{vehicle ... within 10 meters}.  We analyzed all 30 failed tests for S3 and found that they have a vehicle in front of the ego vehicle in the same lane and within 10 meters.  These are false positives, since  the precondition mentions that there should not be any vehicle within 10 meters. We conjecture that the fine-tuned generative model was not able to accurately learn the precise distance relationship; this could be challenging as discerning whether a vehicle in 9.5 or 10.5 meters away is difficult for a human.
 
To substantiate this claim, we conducted a study by adjusting the filtering of data for fine-tuning a LoRA for S3. The original training dataset for S3 includes images with vehicles within 16 to 25 meters range in front of the ego vehicle. We generated a new training dataset by excluding the images with vehicles within 16 to 25 meters range and in front of the ego vehicle. We then used this dataset to fine-tune a LoRA model and generated 1000 tests. Only one out of 1000 tests generated using this LoRA failed the postcondition, and this test also failed the precondition. The percentage of false positives for this configuration is 0.1\% which is a significant reduction from the original S3's 11.3\%.  

From this deeper study of S2 and S3, which is also applicable to S1,
we see the need for further improvements in semantic feature
labeling.   SGSM labeling relies on high-quality scene graphs
and we plan to explore the incorporation of more advanced methods
that can more accurately perform depth estimation, which would
allow \technique to better reflect LC requirements to reduce
false positive rates while enabling fault detection.


\begin{table}[t!]
    \centering
    \begin{tabular}{|c|c|c|} 
    \hline
    Id & Passed Tests & Failed Tests \\
        \hline
        C1 & 
        \begin{minipage}{0.35\textwidth}
            \centering
            \vspace{1pt}
        \includegraphics[width = 0.9\linewidth]{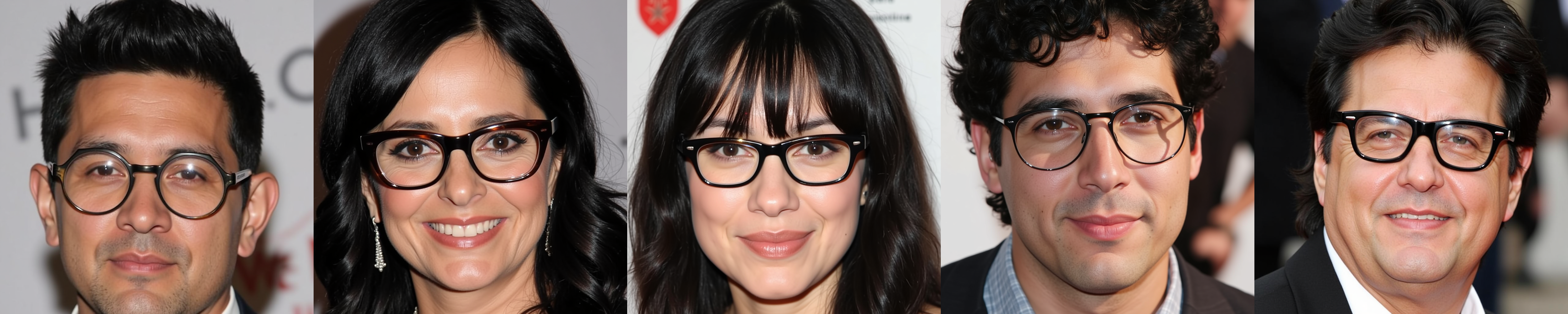} 
        \vspace{1pt}
        \end{minipage} &
        \begin{minipage}{0.35\textwidth}
            \centering
            \vspace{1pt}
        \includegraphics[width=0.9\linewidth]{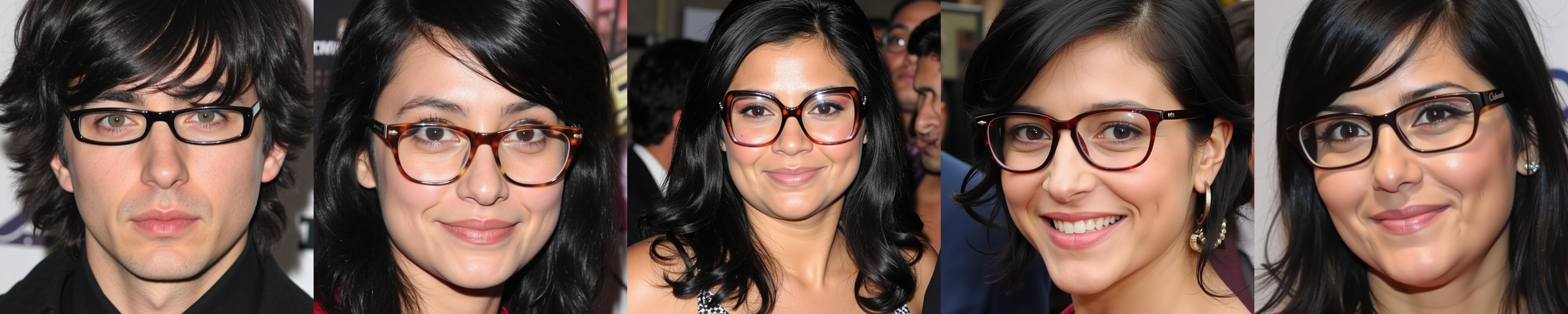} 
        \vspace{1pt}
        \end{minipage} \\ 
        \hline
        
        C4 & 
        \begin{minipage}{0.35\textwidth}
            \centering
            \vspace{1pt}
            \includegraphics[width=0.9\linewidth]{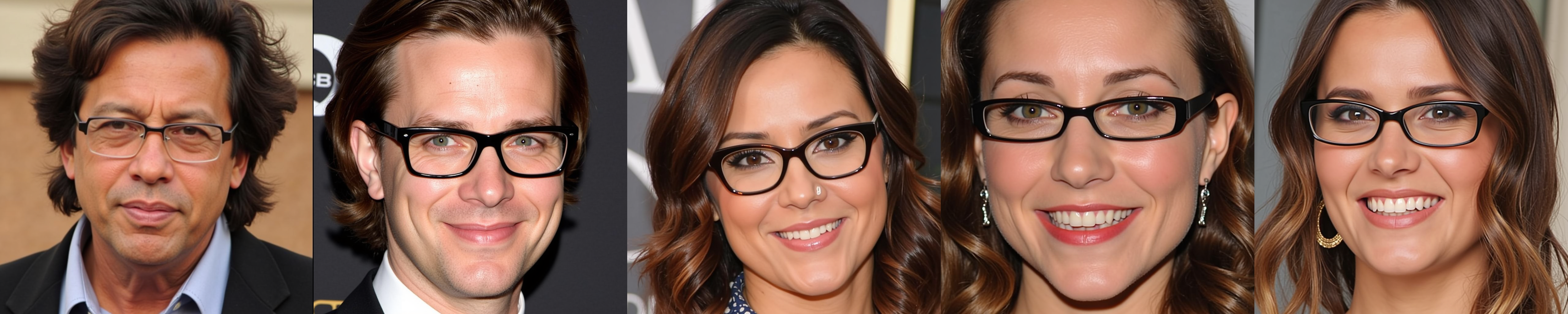}
            \vspace{1pt}
        \end{minipage} &
        \begin{minipage}{0.35\textwidth}
            \centering 
            \vspace{1pt}
        \includegraphics[width=0.9\linewidth]{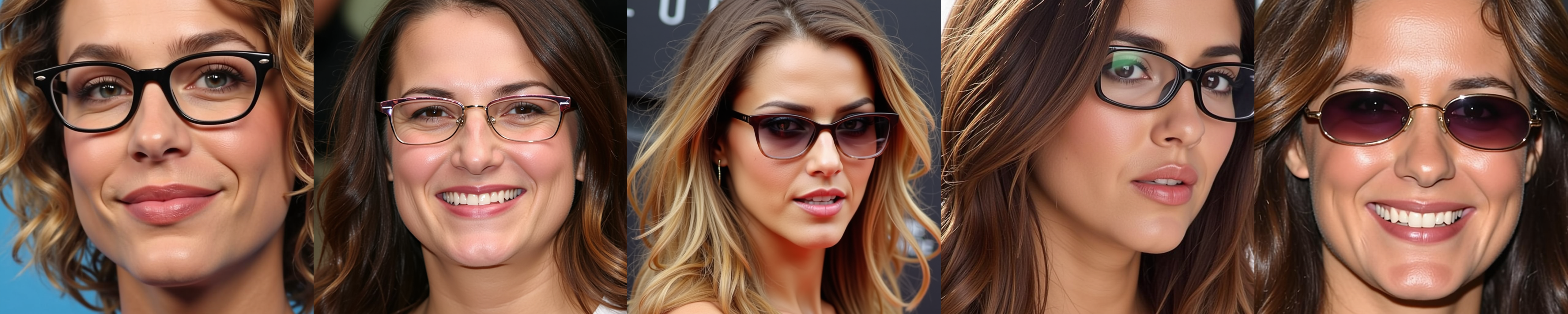} 
        \vspace{1pt}
        \end{minipage} \\
        \hline

        S5 & 
        \begin{minipage}{0.45\textwidth}
            \centering
            \vspace{1pt}
            \includegraphics[width=1\linewidth]{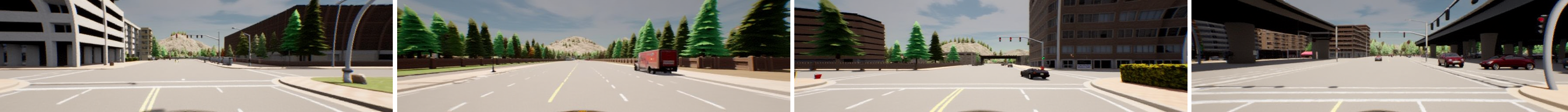}
            \vspace{1pt}
        \end{minipage} &
        \begin{minipage}{0.45\textwidth}
            \centering 
            \vspace{1pt}
        \includegraphics[width=1\linewidth]{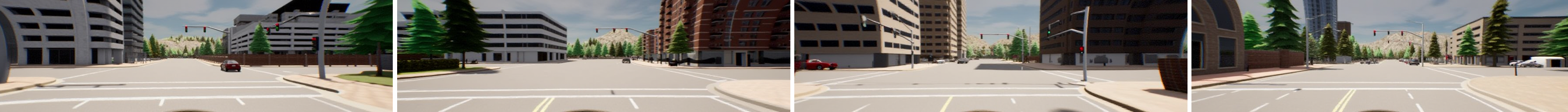} 
        \vspace{1pt}
        \end{minipage} \\
        \hline
        
        S6 & 
        \begin{minipage}{0.45\textwidth}
            \centering
            \vspace{1pt}
            \includegraphics[width=1\linewidth]{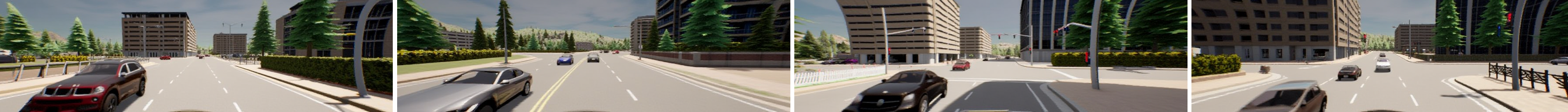}
            \vspace{1pt}
        \end{minipage} &
        \begin{minipage}{0.45\textwidth}
            \centering 
            \vspace{1pt}
        \includegraphics[width=1\linewidth]{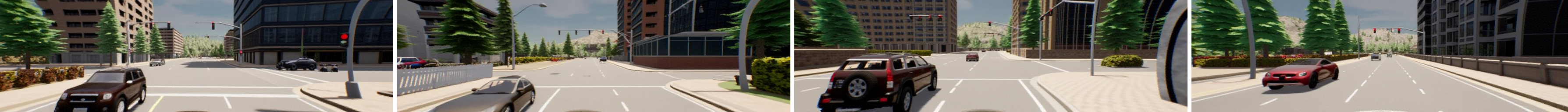} 
        \vspace{1pt}
        \end{minipage} \\
        \hline
    \end{tabular}
    \caption{Random \technique-generated passing and failing test inputs for requirement C1, C4, S5 and S6}
    \label{tab:violationsamples}
\end{table}

\noindent\fbox{
    \parbox{0.97\textwidth}{
       \textbf{\rq{5} Finding: The tests generated by \technique are able to generate failures for 19 of the 21 requirements across our study with false positive rates of 3.3\% on average and 11.5\% in the worst-case.}
}}

\subsubsection{\rq{6}: How effective are \technique-generated tests in revealing a model’s decision behavior?}
\label{sssec:rq6}

Understanding a model's decision-making behavior is crucial for real-world deployment. 
One can view \technique as a means to explore 
model behavior relative to a given set of inputs, described in preconditions, where postconditions are used to define a notion of \textbf{expected} model behavior.
Violations of postconditions indicate unexpected behavior
and the flexibility of \technique allows developers
to easily explore model decision behavior in this way.

To assess this use case of \technique, we  consider three pre-trained Imagenet LCs, described in \S\ref{ssec:lc_selection}, where requirements
I1-I4 in \autoref{tab:requirements} serve to focus the
exploratory process.
In \S\ref{subsec:design}, we described the ImageNet dataset~\cite{deng2009imagenet}, the VQA-based glossary term labeling, how combinations of glossary terms allow for generation of images that are characteristic of zoological taxa, e.g., class, order, or sub-order,
and how expected behavior, expressed in postconditions,
defines a membership test
based on the WordNet taxonomy.
For example, when ``bird'' features are present, the expected ImageNet categories are those that are elements of the Aves taxonomic class.
We trained \technique LoRA for each of I1-I4 from Table~\ref{tab:requirements}.

\begin{figure}[t!]
    \centering
        \includegraphics[width=0.7\linewidth]{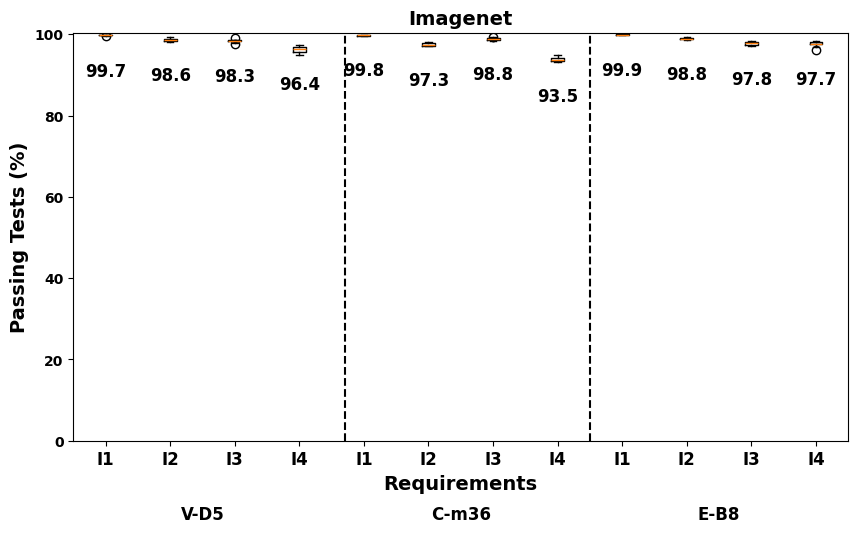}
\caption{Percentage of generated tests for each requirement that passed the post-condition; numerical
values are the means.}
\label{fig:case_study}
\end{figure}

For each precondition, \technique generated 1000 test cases that were executed by each 
of the 3 LCs. The process was followed 10 times, and we report the results as box plots in Figure~\ref{fig:case_study}. 
The pass rate for all three models and four requirements is between 93.5\% and 99.5\%; this indicates that the vast majority of observed model behavior is what is expected.
One reason that pass rates are higher than test accuracy is
that the postcondition
is a weaker test of model behavior than requiring 
a specific test input label -- here the generated label must fall within a set of labels.
This weaker postcondition means, however, that 
violations might represent a more extreme type of unexpected behavior than simply getting the test label wrong.

\begin{figure}[t!]
    \centering
        \includegraphics[width=0.9\linewidth]{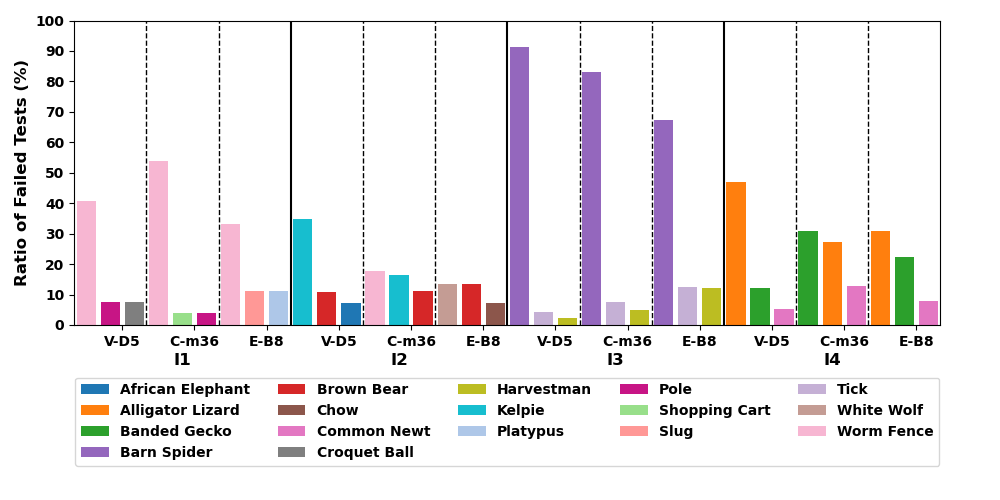}
\caption{Ratio of incorrect predictions by models across predicted classes for each requirement. For each model, the top-3 most frequently predicted incorrect classes with the percentage of total incorrect prediction for a model predicted as those classes are shown individually. }
\label{fig:imagenet_pred}
\end{figure}

\begin{table}[t!]
    \centering
    \begin{tabular}{ccc} 
     &  \textbf{Failure Cases with Model Output} & \textbf{Train Data}\\
        \textbf{I1} & 
        \begin{minipage}{0.4\textwidth}
            \centering
            \vspace{1pt}
        \includegraphics[width = 0.9\linewidth]{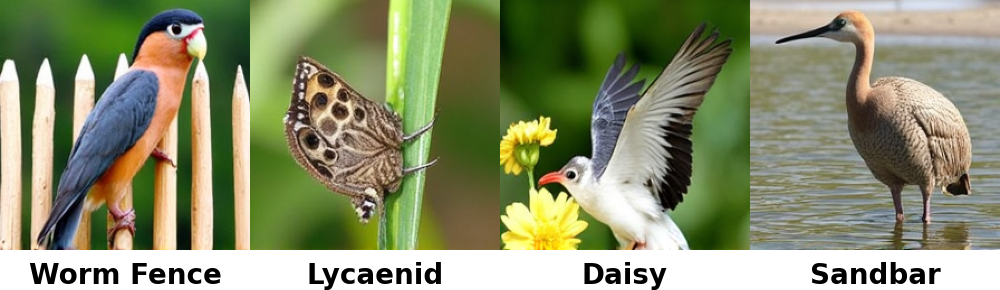} 
        \vspace{1pt}
        \end{minipage} 
        & 
        \begin{minipage}{0.4\textwidth}
            \centering
            \vspace{1pt}
        \includegraphics[width = 0.9\linewidth]{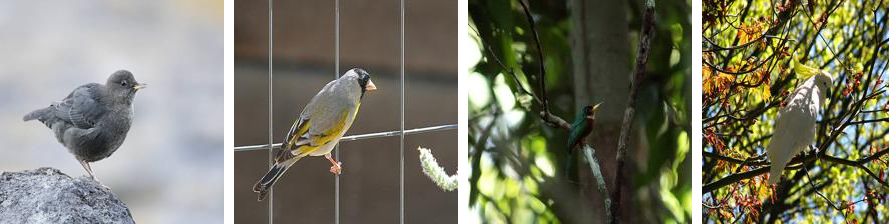} 
        \vspace{1pt}
        \end{minipage} \\ 
        
        \textbf{I2} & 
        \begin{minipage}{0.4\textwidth}
            \centering
            \vspace{1pt}
            \includegraphics[width=0.9\linewidth]{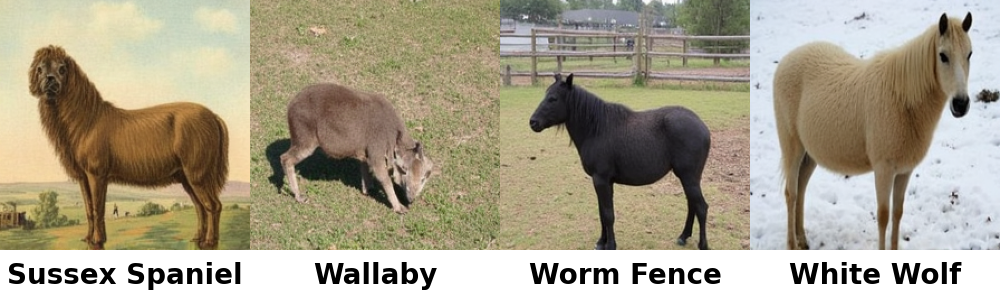}
            \vspace{1pt}
        \end{minipage} & 
        \begin{minipage}{0.4\textwidth}
            \centering
            \vspace{1pt}
        \includegraphics[width = 0.9\linewidth]{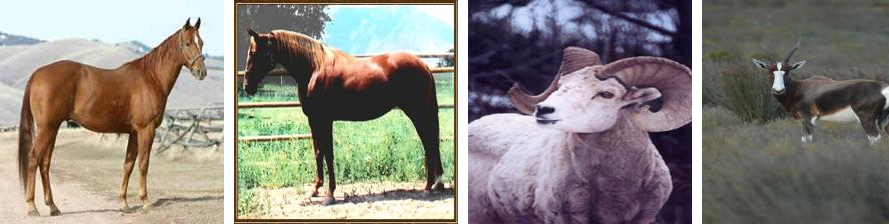} 
        \vspace{1pt}
        \end{minipage} \\

        \textbf{I3} & 
        \begin{minipage}{0.4\textwidth}
            \centering
            \vspace{1pt}
            \includegraphics[width=0.9\linewidth]{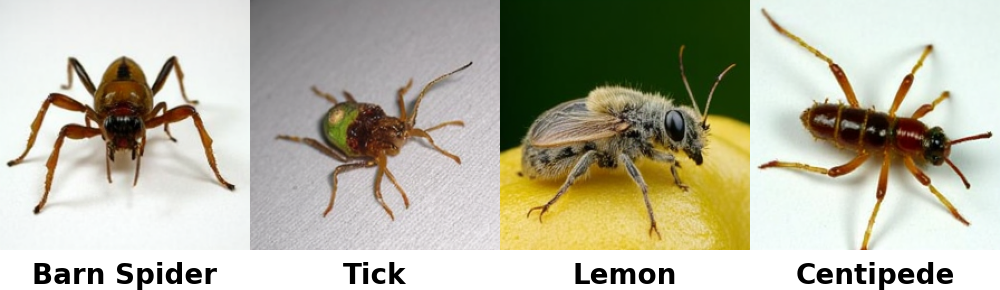}
            \vspace{1pt}
        \end{minipage}  & 
        
        \begin{minipage}{0.4\textwidth}
            \centering
            \vspace{1pt}
            \includegraphics[width=0.9\linewidth]{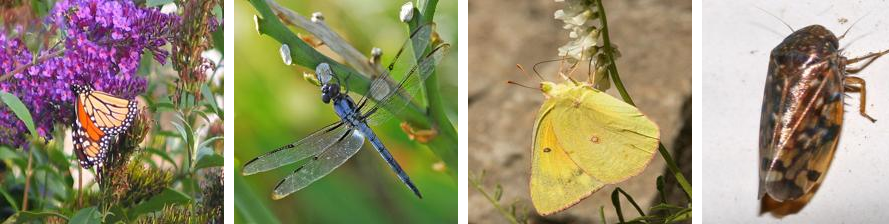}
            \vspace{1pt}
        \end{minipage} \\ 

        \textbf{I4} & 
        \begin{minipage}{0.4\textwidth}
            \centering
            \vspace{1pt}
            \includegraphics[width=0.9\linewidth]{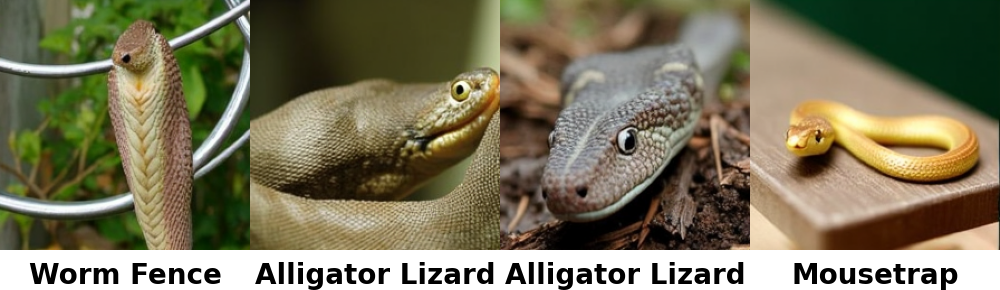}
            \vspace{1pt}
        \end{minipage} & 
        
        \begin{minipage}{0.4\textwidth}
            \centering
            \vspace{1pt}
            \includegraphics[width=0.9\linewidth]{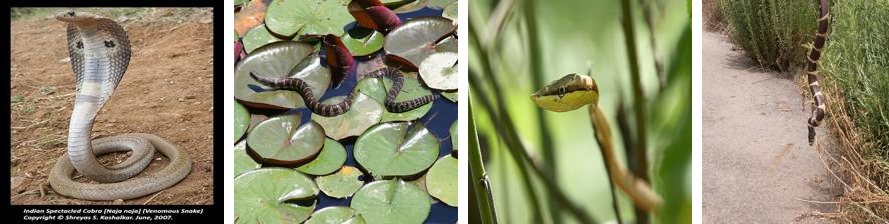}
            \vspace{1pt}
        \end{minipage}

    \end{tabular}
    \caption{Randomly selected test samples with model outcome that led to incorrect predictions and training samples for each requirement.}
    \label{tab:imagenetfailures}
\end{table}

The model behavior can be influenced by many factors, but to demonstrate \technique for this use case we focus our analysis on (a) whether there are issues in the definition of the dataset that
might be lead to unexpected behavior, and (b) whether there are issues in the training process of specific LCs that might lead to unexpected behavior.
We investigated the 1,486 total failures across all three models.
The highest number of failures were observed for requirement I4 with 0.68\% false positive.
Figure~\ref{fig:imagenet_pred} shows the prediction distribution for the failure tests. 
For each model and requirement, the x-axis shows the top 3 most frequent predicted classes that failed. 
In contrast, the y-axis for a class and a model shows the ratio of failed tests specific to that model that were predicted as that class.
According to the figure, the majority of misclassified tests are concentrated within the top three predicted classes, indicating a strong bias in the model’s misclassification patterns.
For instance, more than 50\% of the tests failed by CAformer-M36 model were classified as ``Worm Fence'' for requirement I1.
For requirement I3, all models show a strong bias in their mispredictions toward the class ``Barn Spider'', with over 67\% of their failed tests classified as that category.

To understand the reason behind the dominance of those classes over the failure cases, we manually looked at random samples from incorrect predictions. As shown in Table~\ref{tab:imagenetfailures}, we found that the background of the image influenced the majority of the predictions. Hence, if an animal has an object from a class in its background, the model  may classify the image according to the background. Furthermore, the model often mispredicts when the image contains only a partial view of the animal or when the animal’s body is positioned in a way that resembles another animal or object, for example, predicting a snake as an alligator lizard or a mousetrap, or misclassifying an insect as a barn spider. 

Further investigation of the training images reveals that the Imagenet images contain objects from multiple classes, while they are labeled for a single class only. We conclude that the \technique-generated faults point towards a significant limitation of the Imagenet dataset for single-class classification. Our finding also aligns with prior work~\cite{yun2021re} that acknowledges the limitation and recommends further modification of the Imagenet single labeling to multiclass labels. Our analysis also points towards a key insight that models should be trained to predict multiple classes rather than just a single label. 
While current models attempt to mitigate the ImageNet limitation by providing top-k predictions, they still suffer from the noise introduced during training, especially when images contain multiple objects or only partial views of the primary subject, such as an animal. As a result, top-k predictions do not fully resolve the issues caused by single-label supervision. 

Furthermore, \technique provides valuable insights into a model’s failure modes. 
For instance, as shown in Figure~\ref{fig:imagenet_pred}, requirement I2 highlights different failure patterns among the models. 
The failed tests for EfficientNet-B8 are predominantly misclassified as ``White Wolf'', whereas this class does not appear among the top three mispredicted classes for CAformer-M36 or VOLO-D5.
These findings provide concrete, model-specific insights that can inform actionable recommendations for improving the ImageNet dataset and the models trained on it.
Our technique therefore not only diagnoses distinct failure modes across models but also guides targeted strategies for refining both the ImageNet dataset and model training practices.

\noindent\fbox{
    \parbox{0.97\textwidth}{
       \textbf{RQ6 Finding: Applying \technique with VQA-based glossary term labeling to test 3 ImageNet LCs for a collection of 4 semantic-feature functional requirements yielded insights into LC failure modes with very low false positive rates.}
}}

\subsection{Threats to Validity}
To mitigate threats to internal validity, wherever possible, we use existing infrastructure, like FLUX.1-dev~\cite{fluxmodel} and AI Toolkit\cite{aitoolkitLoRA}, which are in wide-spread use and
very actively maintained.  Nevertheless we are combining these methods along with \technique
specific methods, such as a scene-graph based autolabeler, to create novel capabilities.
Ideally, we would be able to formulate precise correctness specifications for generated images
and check those on the outputs of our approach, but this is precisely the problem we are targeting -- generating inputs when there are no precise specifications.   To compensate for this lack of
precise internal checking, we performed significant human analysis of random samples from all of
the generative models discussed in \S~\ref{sec:evaluation}.  This involved co-authors performing
independent assessments and using their combined results to validate the generated tests.

We chose four datasets that reflect very different domains in order to provide a degree
of generalizability in our findings.  More datasets would add value and we plan to significantly
expand the variety of SGSM-like datasets in future work.   One reason for this is that the
driving domain has a growing body of safety specifications that can be leveraged and our aim
is to generalize to such specifications.   While we considered two groups of semantic input-output relation and two groups of semantic input-output robustness requirements, further exploration of these requirement types would improve generalizability. 

Wherever possible, we chose standard metrics used elsewhere in the machine learning and
software testing community.  For example, we measured fault detection rate and then analyzed
failing tests to estimate the false positive rate in RQ5, used KID in RQ2, and JS in RQ3.
Our prediction match metric is reminiscent of the fault detection metric, but is applied
to the input rather than the output of the model.   While a broader range of metrics might
add value, the chosen metrics provide information that directly relates to the research questions.   Our qualitative evaluation on random samples complements, and is consistent
with, the metrics reported and we share a richer set of samples in our open-source project repository~\cite{rbt4dnn}.

\section{Conclusions and Future Work}
\label{sec:conclusion}
\technique is the first test generation technique for neural networks that drives test input generation based on requirement preconditions expressed in a semantic feature space.  This allows the network output for those test inputs to be checked against
postconditions that are tailored to the precondition.   
Our experimental evaluation of \technique  
demonstrates that it is capable of generating
test inputs that are consistent with preconditions,
that are diverse and realistic relative to training datasets, and
that can reveal faults and unobserved behavior in well-trained LCs.   

While this paper presents a first step towards leveraging feature-based functional
requirements for validation of LCs, there are many fruitful directions for future work.   
There is potential for improving training and fine-tuning of LDMs that could increase fault-detection and lower false positive rates.  Further, adapting techniques for  systematic latent space coverage, e.g.,~\cite{dola2024cit4dnn}, would allow \technique to provide evidence of precondition coverage when faults cannot be revealed.

\section{Data Availability}
\label{sec:data_availability}
The Pytorch implementation of RBT4DNN framework and its data can be found here: \url{https://github.com/less-lab-uva/RBT4DNN} 

\section{Acknowledgments}
This research is based in part upon work supported by NSF awards 2129824, 2217071, and 2312487 and by ARO grant number W911NF-24-1-0089.   The authors acknowledge Research Computing at The University of Virginia for providing access to computational resources.

\bibliographystyle{IEEEtran.bst}
\bibliography{IEEEfull, main}

\end{document}